\documentclass[aps,pre,twocolumn]{revtex4}

\usepackage{amsmath,bm,epsfig}
\def \ed {\end{document}}
\def\Fbox#1{\vskip1ex\hbox to 8.5cm{\hfil\fboxsep0.3cm\fbox{%
  \parbox{8.0cm}{#1}}\hfil}\vskip1ex\noindent}  %%  {TEXT} in BOX

\let\*\cdot
\def\<{\left\langle} \def\>{\right\rangle} \def\({\left(} \def\){\right)}
 \let\~\widetilde \let\^\widehat 
   
  \def\1{\bm1} 
 
\newcommand{\B}[1]{{\bm{#1}}}%% Bold Roman & Greek Lower & Upper Case
\newcommand{\eq}[1]{(\ref{#1})}%%  requires \eq{label}
\newcommand{\Eq}[1]{Eq.~(\ref{#1})}%%  requires \eq{label}
\newcommand{\Eqs}[1]{Eqs.~(\ref{#1})}%%  requires \eq{label}
\def\<{\left\langle} \def\>{\right\rangle} \def\({\left(} \def\){\right)}
 \let\~\widetilde
\def\Ret {\mbox{Re}_\tau}
\renewcommand{\sb}[1]{_{\text {#1}}} %% sub- for lower case
\newcommand{\Sub}[1]{_{_{\text {#1}}}} %% Sub- for Upper case
\begin{document}

\title{The State of the Art in Hydrodynamic Turbulence: Past Successes and Future Challenges}

\author{Itamar Procaccia$^*$ and K.R. Sreenivasan$^\dag$}
\affiliation{$^*$Department  of Chemical Physics, The Weizmann Institute
of Science, Rehovot 76100, Israel\\
$^\dag$ International Centre for Theoretical Physics, Strada
Costiera 11, 34014 Trieste, Italy}
\begin{abstract}
We present a personal view of the state of the art in turbulence
research. We summarize first the main achievements in the recent
past, and then point ahead to the main challenges that remain for
experimental and theoretical efforts. \end{abstract} \maketitle
\tableofcontents
\section{Introduction}\
\label{intro} ``The problem of turbulence" is often hailed as one of
the last open problems of classical physics. In fact, there is no
single ``problem of turbulence"; rather, there are many
inter-related problems, some of which had seen significant progress
in recent years, and some are still open and inviting further
research. The aim of this short review is to explain where
fundamental progress has been made and where, in the opinion of the
present writers, there are opportunities for further research.

There are many ways to set a fluid into turbulent motion. Examples
include creating a large pressure gradient in a channel or a pipe,
pulling a grid through a fluid, moving one or more boundaries to
create a high shear and forcing a high thermal gradient. Customarily
the vigor of forcing is measured by the Reynolds number $Re$,
defined as $Re\equiv U L/\nu$ where $L$ is the scale of the forcing,
$U$ is the characteristic velocity of the fluid at that same scale,
and $\nu$ is the kinematic viscosity. The higher the Reynolds number
the larger is the range of scales involved in the turbulent motion,
roughly from the scale $L$ itself (known as the ``outer" or
``integral" scale) down to the so-called ``viscous" scale $\eta$
which decreases as Re$^{-3/4}$ \cite{monin}. For large $Re$ a
turbulent flow exhibits an erratic dependence of the velocity field
on the position in the fluid and on time. For that reason it is
universally accepted that a statistical description of turbulence is
called for, such that the objects of interest are almost invariably
mean quantities (over time, space or an ensemble, depending on the
application), fluctuations about the mean quantities, and
correlation functions defined by these fluctuations; precise
definitions will be given below. Thus, the crucial scientific
questions deal typically with the universality of the statistical
objects, universality with respect of the change of the fluid, or
universality with respect to the change of forcing mechanisms. We
will see that this universality issue binds together the various
aspects of turbulence to be discussed below into a common
quest---the quest for understanding those aspects of the phenomenon
that transcend particular examples. We will strive to underline
instances when this quest has been successful and when doubts
remain.

The structure of this review is as follows: in Sect.\
\ref{anomalous} we discuss the statistical theory of homogeneous
and isotropic turbulence and focus on the anomalous scaling
exponents of correlation functions. For a part of the community this
represented {\em the} important open problem in turbulence, and indeed great
progress had been achieved here. In Sect.\ \ref{anisotropy} we
address homogeneous but anisotropic turbulence and present recent
progress in understanding how to extract information about the
isotropic statistical objects, and how to characterize the anisotropic
contributions. Section \ref{bounded} deals with wall-bounded
turbulence where both isotropy and homogeneity are lost (being
actually the norm in practice, rather than the exception). We focus
on the controversial issue of the log law versus power laws,
clarifying the scaling assumptions underlying each of these
approaches and replacing them by a universal scaling function; we
show that this achieves an excellent modeling of channel or pipe
flows. In Sect. \ref{drag} we consider turbulence with additives
(like polymers or bubbles) and review the progress in understanding
drag reduction by such additives. Section \ref{convection} discusses
problems in thermal convection, with emphasis on recent work.
Finally, Sect.\ \ref{superfluid} provides a selective account of the
problems that have come to the fore in superfluid turbulence,
sometimes bearing directly on its classical counterpart. The article
concludes with a summary of outlook.

\section{Anomalous Scaling in Homogeneous and Isotropic Turbulence}
\label{anomalous} A riddle of central interest for more than half a
century to the theorist and the experimentalist alike concerns the
numerical values of the scaling exponents that characterize the
correlation and structure functions in homogeneous and isotropic
turbulence. Before stating the problem one should note that strictly
homogeneous and isotropic state of a turbulent flow is not
achievable in experiments; typically the same forcing mechanism that
creates the turbulent flow is also responsible for breaking
homogeneity or isotropy. Nevertheless, some reasonable
approximations have been attained. To get a closer approximation,
one has to resort to numerical simulations. For a long time, the
Reynolds number of simulations was limited by numerical resolution
and by storage capabilities, but this situation has improved
tremendously in the past few years. Indeed, as an idealized state of
turbulence which incorporates the essentials of the nonlinear
transfer of energy among scales, homogeneous and isotropic
turbulence has gained a time-honored status in the history of
turbulence research.

Consider then the velocity field $\B u(\B r,t)$ which satisfies the
Navier-Stokes equations
\begin{equation}
\frac{\partial \B u}{\partial t} + \B u \cdot \B \nabla \B u =-\B \nabla p +\nu \nabla^2 \B u + \B f \ , \label{NS}
\end{equation}
where $p$ is the pressure and $\B f$ the (isotropic and homogeneous)
forcing that creates the (isotropic and homogenous) turbulent flow.
Defining by $\langle \dots \rangle$ an average over time, we realize
that $\langle \B u \rangle=0$ everywhere in this flow. On the other
hand, correlations of $\B u$ are of interest, and we define the
so-called ``unfused" $n$th order correlation function $\B T_n$ as
\begin{equation}
\B T_n (\B r_1, t_1,\B r_2, t_2,\dots \B r_n,t_n)\! \equiv\! \langle
\B u(\B r_1,t_1)  \B u(\B r_2,t_2) \dots  \B u(\B r_n,t_n) \rangle \
. \label{defFn}
\end{equation}
When all the times $t_i$ are the same, $t_i=t$, we get the
equal-time correlation function $\B F_n(\B r_1 ,\B r_2,\dots \B
r_n)$ which, for a forcing that is stationary in time, is a
time-independent function of the $n(n-1)/2$ distances between the
points of measurements, due to homogeneity. An even more contracted
object is the so called ``longitudinal structure function" $S_n$,
\begin{equation}
S_n(R) \equiv \langle\{[ \B u (\B r+\B R,t)-\B u (\B r,t)]\cdot \B R/R\}^n\rangle \ , \label{defSn}
\end{equation}
which can be obtained by sums and differences of correlation
functions $\B F_n$, together with some fusion of coordinates \cite{96LP}. On the
basis of evidence from experiments and simulations, it has been
stipulated (although never proven) that $S_n$ is a homogeneous
function of its arguments when the distance $R$ is within the
so-called ``inertial range" $\eta\ll R \ll L$ in the sense that
\begin{equation}
S_n(\lambda R) = \lambda^{\zeta_n} S_n(R) \ . \label{defzetan}
\end{equation}
The central question concerns the numerical values of the ``scaling
exponents" $\zeta_n$ and their universality with respect to the
nature of the forcing $f$. This question poses serious difficulties
since it is impossible to derive a closed form theory for a given
order structure function $S_n$, since any such theory involves
higher order unfused correlation functions with integrations over
the time variable \cite{98BLPP,00LP}.

A closely related question with lesser theoretical difficulties pertains
to other fields that couple to the velocity field, with the
``passive scalar" case drawing most attention during the nineties. A
passive scaler $\phi(\B r,t)$ is a field that is advected by a
turbulent field which itself is unaffected by it. For example,
\begin{equation}
\frac{\partial \phi}{\partial t} + \B u \cdot \B \nabla \phi =
\kappa \nabla^2 \phi + f \ .\label{passive}
\end{equation}
If $\B u$ and $f$ are homogeneous and isotropic, and Re$\to \infty$
and $\kappa\to 0$, the structure functions ${\cal S}_n\equiv \langle
[\phi(\B r+\B R)-\phi(\B r)]^n\rangle$ are stipulated to be
homogeneous functions of the their arguments with scaling exponents
$\xi_n$.

Dimensional considerations predict $\zeta_n=\xi_n=n/3$, with
$\zeta_3=1$ being an exact result from fluid mechanics, going back
to Kolmogorov \cite{41Kol}. Experimental and simulations data
deviated from these predictions (except, of course, for $n=3$), and a
hot pursuit for an example where these exponents could be calculated
theoretically was inevitable. The first example that yielded to
analysis was the Kraichnan model \cite{68Kra}, in which $\B u$ is not
a generic velocity field, but rather a random Gaussian field whose
second order structure function scales with a scaling exponent
$\zeta_2$ as in Eq.\ (\ref{defzetan}), but is $\delta$-correlated in
time. This feature of the advecting field leads to a great
theoretical simplification, not as much as to provide a closed form
theory for ${\cal S}_n$, but enough to allow a derivation of a
differential equation for the simultaneous $2n$-th order correlation
function ${\cal F}_{2n} =\langle \phi(\B r_1)\dots \phi(\B
r_{2n})\rangle$, having the symbolic form \cite{68Kra}
\begin{equation}
{\cal O} {\cal F}_{2n} = RHS ({\cal F}_{2n-2}) \ . \label{diffeq}
\end{equation}
Guessing the scaling exponent of $F_{2n}$ by power counting and
balancing the LHS against the RHS yields dimensional scaling
estimates which, in this case, are $\xi_{2n} =(2-\zeta_2)n$. The
crucial observation, however, is that the differential equations
(\ref{diffeq}) possess homogeneous solutions of the equation ${\cal
O} {\cal F}_{2n}=0$ \cite{95GK,95CFKL}. These ``zero modes" are homogeneous functions
of their arguments but their exponent cannot be guessed from power
counting; the scaling exponents are anomalous---i.e., $\xi_{2n} <
(2-\zeta_2)n$---and therefore dominant at small scales. The
exponents could be computed in perturbation theory around $\zeta_2 =
0$, demonstrating for the first time that dimensional scaling
exponents are not the solution to the problem.

An appealing interpretation of the physical mechanism for anomalous
exponents of the Kraichnan model was presented in the framework of
the Lagrangian formulation \cite{98GPZ}. In this formulation an $n$th order
correlation function results from averaging over all the Lagrangian
trajectories of groups of $n$ fluid points that started somewhere at
$t=-\infty$ and ended their trajectories at points $\B r_1\dots \B
r_n$ at time $t=0$. Analyzing this dynamics it turned out that the
Richardson diffusion of these groups did not contribute anything to
the anomalous scaling. Rather, it is the dynamics of the shapes
(triangles for 3 points, tetrahedra for 4 points etc) that is
responsible for the anomaly. In fact, the anomalous scaling
exponents could be related to eigenvalues of operators made from the
shape-to-shape transition probability \cite{01AP}. The zero modes discussed
above are {\em distributions over the space of shapes} that remain
invariant to the dynamics \cite{01CV}. It appears that these findings of the
importance of shapes rather than scales in determining anomalous
exponents is a new contribution to the plethora of anomalous
exponents in field theory, and it would be surprising if other
examples where shapes rather than scales are crucial will not pop-up
in other corners of field theory, classical or quantum-mechanical.

The finding of distributions that remain invariant to the dynamics
meant that also in the Eulerian frame there must be such
distributions, since the change from Lagrangian to Eulerian is just
a smooth change of coordinates. Indeed this was the case, and this
was the clue how to generalize the results of the non-generic
Kraichnan model to the generic case Eq. (\ref{passive}) with a generic
velocity field that stems from the Navier-Stokes equations. The
central comment is that the decaying passive scalar problem, i.e. Eq
(\ref{passive}) with $f=0$ is a linear problem for which one can
always define a propagator from ${\cal F}_n$ at $t=0$ (i.e. $\langle
\phi(\B r_1,t=0)\dots \phi(\B r_n,t=0)\rangle$ to the same object at
time $t$ (note that for the decaying problem this is no longer a
stationary quantity) \cite{01ABCPM}. This propagator possesses eigenfunctions of
eigenvalue 1 which are homogeneous functions of their arguments,
characterized by anomalous exponents. These are the analogs of the
zero modes of the Kraichnan model, and they are responsible for the
anomalous exponents in the generic case \cite{02CTP,03CPP}. Thus the general statement
that can be made is that the anomaly in the case of passive scalar,
generic or not, is due to the existence of ``statistically preserved
structures"; these can increases or decrease in every single
experiment, but remain invariant on the average. This is a novel
notion that pertains to nonlinear nonequilibrium systems without a
known analog in equilibrium statistical physics.

At present it is still unclear whether the insight gained from linear models might have
direct relevance to the nonlinear problem itself. Some positive indications in this direction
can be found in \cite{06ABBP}, but much more needs to be done here before conclusions can be drawn.

\section{Statistical Theory of Anisotropic Homogeneous Turbulence}
\label{anisotropy}

As mentioned above, the very same agents that force turbulence tend
to destroy also homogeneity and isotropy. In this section we are
concerned about the loss of isotropy and review the extensive work
that has been done to come to grips with this issue in a systematic
fashion. Since this subject has been reviewed extensively
\cite{05BP}, we limit this section to only a few essential
comments.

The need for rethinking the issue of loss of isotropy was underlined
by the appearance of several papers where anisotropic flows were
analyzed disregarding anisotropy, and exponents were extracted from
data assuming that the inertial range scales were isotropic. The
results were confusing: scaling exponents varied from experiment to
experiment, and from one position in the flow to another. If this were
indeed the case, the notion of universality in turbulence would fail
irreversibly. In fact, it can now be shown that all these worrisome
results can be attributed to anisotropic contributions in the
inertial range, as explained below.

The basic idea in dealing with anisotropy is that the equations of
fluid mechanics are invariant to all rotations. Of course, these
equations are also nonlinear, and therefore one cannot foliate them
into the sectors of the SO(3) symmetry group. The equations for
correlation functions are, however, linear (though forming an
infinite hierarchy). Thus by expanding the correlation functions in
the irreducible representations of the symmetry group, one gets a
set of equations that are valid sector by sector  \cite{99ALP}. The irreducible
representations of the SO(3) symmetry group are organized by two
quantum numbers $j,m$ with $j=0,1,2,\dots$ and $m=-j,-j+1,\dots j$.
It turns out that the $m$ components are mixed by the equations of
motion, but the $j$ components are not. Accordingly one can show
that an $n$-point correlation function admits and expansion
\begin{equation}
\B F_n(\B r_1 , \B r_2,\dots \B r_n)=\sum_{qjm}
A_{qjm}(r_1,r_2,\dots, r_n) \B B_{qjm}({\bf  \hat r_1}, {\bf \hat
r_2}\dots {\bf \hat  r_n}) \ , \label{so3}
\end{equation}
where ${\bf \hat r}$ is a unit vector in the direction of $\B r$,
and $A_{qjm}$ is a homogeneous function of the scalar $r_1\dots
r_n$,
\begin{equation}
A_{qjm}(\lambda r_1,\lambda r_2,\dots, \lambda r_n)=
\lambda^{\zeta_n^{(j)}}A_{qjm}(r_1,r_2,\dots, r_n) \ .
\end{equation}
Here $\zeta_n^{(j)}$ is the scaling exponent characterizing the
$j$-sector of the symmetry group for the $n$th order correlation
function. $\B B_{qjm}( {\bf \hat r_1}, {\bf \hat r_2}\dots {\bf \hat
r_n})$ are the $n$-rank tensorial irreducible representations of the
SO(3) symmetry group, and the index $q$ in Eq.\ (\ref{so3}) is due
to the fact that higher order tensors have more than one irreducible
representation with the same $jm$  \cite{99ALP}.

It was shown that this property of the $n$-th order correlation
functions is inherited by the structure functions as well \cite{98ADKLS}. Since
these are scalar functions of a vector argument they get expanded in
standard spherical harmonics $\phi_{jm}({\bf \hat R})$
\begin{equation}
S_n(\B R) = \sum_{jm} a_{jm}(r)\phi_{jm}({\bf \hat R}) \ ,
\end{equation}
with
\begin{equation}
a_{jm}(\lambda r) =\lambda^{\zeta_n^{(j)}} a_{jm}(r) \ .
\end{equation}
The main issue for research was the numerical values of this
plethora of scaling exponents. The subject was reviewed in depth in
\cite{05BP}, and we therefore limit our comments here to just the bare
essentials.

Of considerable help in organizing the scaling exponents in the
various sectors of the symmetry group where the Kraichnan model and
related models (like the passive vector model with pressure), where
the exponents could be computed analytically in the Eulerian frame
in any sector of the symmetry group. The central quantitative result
of the Eulerian calculation is the expression for the scaling
exponent $\xi^{(n)}_{j}$ which is associated with the scaling
behavior of the $n$-order correlation function (or structure
function) of the scalar field in the $j$-th sector of the symmetry
group. In other words, this is the scaling exponent of the
projection of the correlation function on the $j$-th irreducible
representation of the SO($d$) symmetry group, with $n$ and $j$
taking on even values only, $n=0,2, \dots$ and $j=0,2,\dots$ \cite{00ALPP}:
\begin{equation}
\xi^{(n)}_{j}= n-\epsilon\Big[\frac{n(n+d)}{2(d+2)}
-\frac{(d+1)j(j+d-2)}{2(d+2)(d-1)}\Big] +O(\epsilon^2) \ .
\label{eq:perturbative}
\end{equation}
The result is valid for any even $j\le n$, and to $O(\epsilon)$. In
the isotropic sector ($j=0$) we recover the  result of \cite{95GK}.
It is noteworthy that for higher values of $j$ the discrete spectrum
is a strictly increasing function of $j$. This is important, since
it shows that for diminishing scales the higher order scaling
exponents become irrelevant, and for sufficiently small scales only
the isotropic contribution survives. As the scaling exponent appear
in power laws of the type $(r/\Lambda)^\xi$, with $\Lambda$ being
some typical outer scale and $r \ll \Lambda$, the larger is the
exponent, the faster is the decay of the contribution as the scale
$r$ diminishes. This is precisely how the isotropization of the
small scales takes place, and the higher order exponents describe
the rate of isotropization. Nevertheless for intermediate scales, or
for finite values of the Reynolds and Peclet numbers, the lower
lying scaling exponents will appear in all the measured quantities,
and understanding their role and disentangling the various
contributions cannot be avoided.

For Navier-Stokes turbulence the exponents cannot be computed
analytically, but the results that are obtained from analyzing both
experiments \cite{98ADKLS} and simulations \cite{99ABMP} indicate that the picture obtained for
the Kraichnan model repeats. The isotropic sector is always leading
(in the sense that scaling exponents belonging to higher sector are
numerically larger). There is growing evidence of universality of
scaling exponents in all the sectors, but this issue is far from
being settled, and more experiments and simulations are necessary to
provide decisive evidence. It is noteworthy that the issue of
universality of the exponents in the isotropic sector is here
expanded many-fold into all the sectors of the symmetry group, and
this is certainly worth further study.

%%%%%%%%%%%%%%%%%%%%%%%%%%%%%%
\section{Wall-Bounded Turbulence}
\label{bounded}

Turbulent flows of highest relevance for engineering application
possess neither isotropy nor homogeneity. For example, turbulent
flows in channels and pipes are strongly anisotropic and
inhomogeneous; indeed, in a stationary plane channel flow with a
constant pressure gradient $p'\equiv -\partial p/\partial x$ the
only component of the mean velocity $\B V$, the streamwise component
$V_x\equiv V$, depends strongly on the wall normal direction $z$; so
do the derivatives of $V_x$ with respect to $z$ and the second order
quantities such as mean-square-fluctuations. A long-standing
challenge for engineers is the description of the {\bf profiles} of
the mean velocity and second order fluctuations throughout the
channel or pipe at relatively high but finite Reynolds numbers.

To understand the issue, focus on a channel of width 2$L$ between
its parallel walls, where the incompressible fluid velocity $\bm
U(\bm r,t)$ is decomposed into its mean (i.e., average over time)
and a fluctuating part
\begin{equation}
\bm U(\bm r,t) = \bm{V}(\bm r)  + \bm u(\bm r,t) \ , \ \bm V(\bm r)
\equiv \langle \bm U(\bm r,t) \rangle \ .
\end{equation}
Near the wall, the mean velocity profiles for different Reynolds
numbers exhibit (to the lowest order) data collapse once presented
in wall units, where the Reynolds number $\Ret$, the normalized
distance from the wall $z^+$ and the normalized mean velocity
$V^+(z^+)$ are defined (for channels) by
$$
\Ret \equiv {L\sqrt{\mathstrut p' L}}/{\nu}\ , \  z^+
\equiv {z \Ret }/{L} \ , \  V^+ \equiv
{V}/{\sqrt{\mathstrut p'L}} \ .
$$

The classical theory of Prandtl and von-K\`arm\`an for infinitely
large $\Ret$ is based on dimensional reasoning and on the assumption
that {\em the single characteristic scale in the problem is
proportional to the distance from the (nearest) wall} (and see below
for details). It leads to the celebrated von-K\`arm\`an
log-law~\cite{monin}
\begin{equation}
V^+(z^+) =  \kappa^{-1}\ln (z^+)  +B \,, \label{loglaw}
\end{equation}
which serves as a basis for the parametrization of turbulent flows
near a wall in many engineering applications. On the face of it this
law agrees with the data (see, e.g. Fig.~\ref{profiles}) for
relatively large $z^+$, say for $z^+>100$, giving $\kappa\sim 0.4$
and $B\sim 5$. The range of validly of the log-law is definitely
restricted by the requirement $\zeta\ll 1$, where  $\zeta \equiv
z/L$ (channel) or $\zeta\equiv r/R$ (pipe of radius $R$).  For
$\zeta\sim 1$ the global geometry becomes important leading to
unavoidable deviations of $V^+ (\zeta)$ from the
log-law~(\ref{loglaw}), known as {\it the wake}.

%%%%%%%%%%%%%%%%%% FIG 1 %%%%%%%%%%%%%%%%%
  \begin{figure*}
\includegraphics[width=0.49\textwidth]{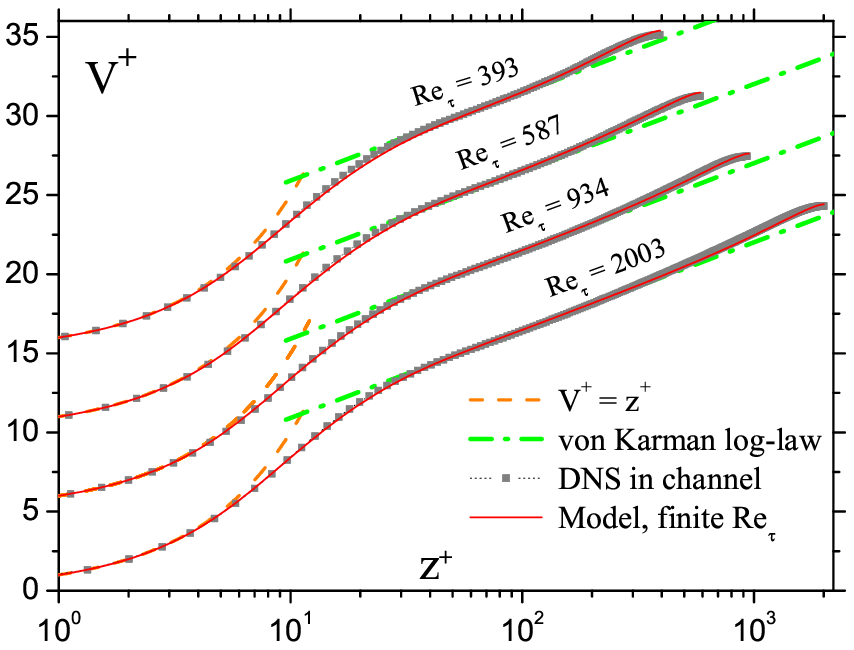}~~~~
\includegraphics[width=0.49 \textwidth]{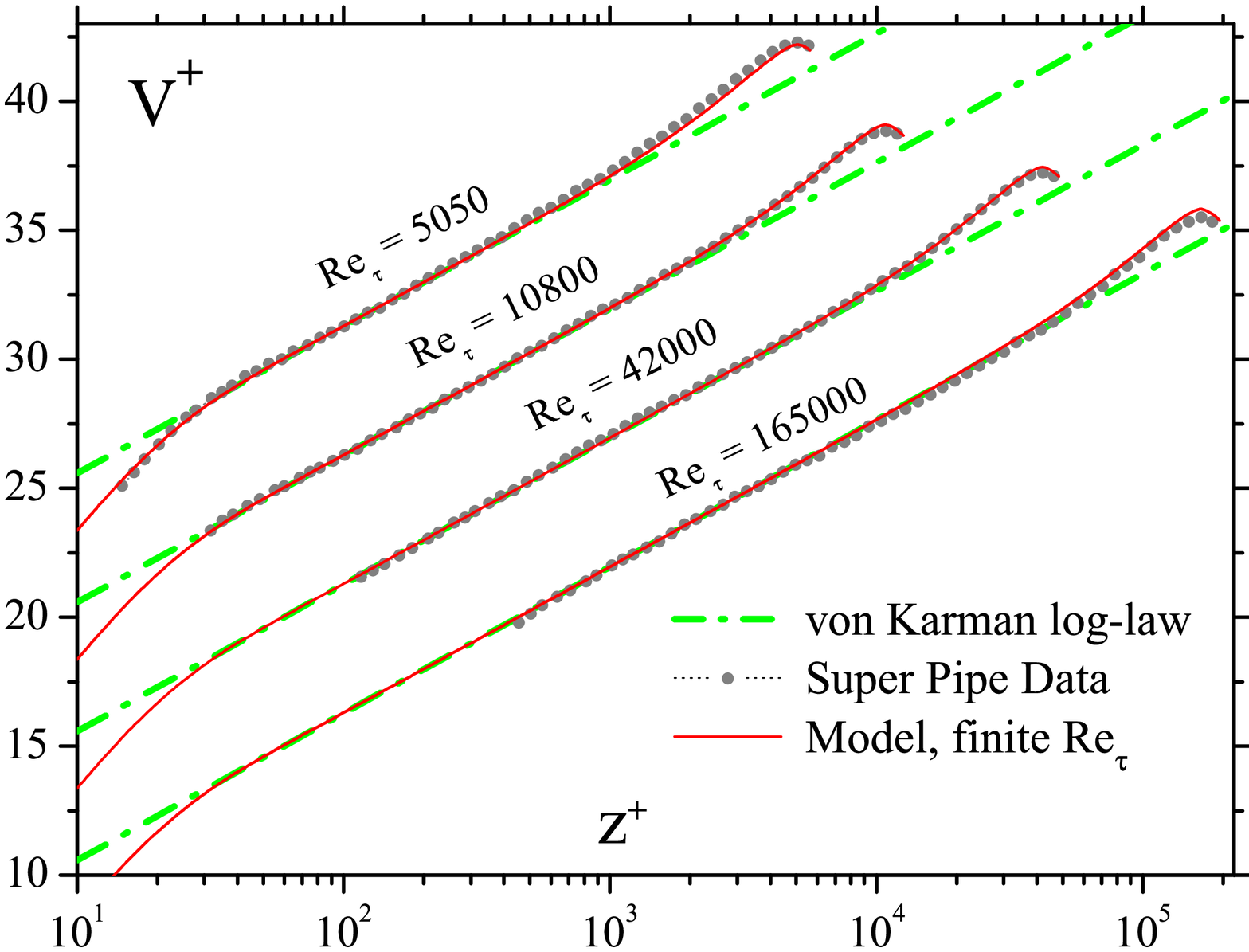}
\caption{Color online. Comparison of the theoretical mean velocity
profiles (red solid lines) at different values of $\Ret$ with the
DNS data  for the channel flow~\cite{Moser,DNS} (Left panel,  grey
squares; model with $\ell\sb{buf}=49,\ \kappa=0.415, \ \ell\sb
s=0.311$) and with the experimental Super-Pipe data~\cite{princeton}
(Right panel, grey circles; model with $\ell\sb{buf}=46,\
\kappa=0.405, \ \ell\sb s=0.275$). In orange dashed line we plot the
viscous solution $V^+=z^+$. In green dashed dotted line we present
the von-K\`arm\`an log-law. Note that the theoretical predictions
with three $\Ret$-independent parameters fits the data throughout
the channel and pipe, from the viscous scale, through the buffer
layer, the log-layer and the wake. For clarity the consequent plots
are shifted vertically on five units.} \label{profiles}
\end{figure*}
%%%%%%%%%%%%%%%%%%%%%%%%%%%%%%%
The problem is that for finite $\Ret$ the corrections to the log-law
(\ref{loglaw}) are in powers of $\varepsilon\equiv 1/\ln
\Ret$~\cite{93Bar} and definitely cannot be neglected for the
currently largest available direct numerical simulation (DNS) of
channel flows ($\Ret=2003$~\cite{Moser,DNS} or $\varepsilon \approx
0.13$). Even for $\Ret$  approaching $500,000$ as in the Princeton
Superpipe experiment~\cite{princeton}, $\varepsilon\approx 0.08$.
This opens a Pandora box with various possibilities to revise the
log-law~(\ref{loglaw}) and to replace it, as was suggested
in~\cite{93Bar}, by a power law
\begin{equation}\label{BC}
V^+(z^+) = C(\Ret) (z^+)^{\gamma(\Ret)}\ .
\end{equation}
Here both the coefficients $C(\Ret) $ and the exponents
$\gamma(\Ret)$ were represented as asymptotic series expansions in
$\varepsilon$. The relative complexity of this proposition compared
to the simplicity of Eq.~(\ref{loglaw}) resulted in a less than
enthusiastic response in the fluid mechanics community \cite{98SZ},
leading to a rather fierce controversy between the log-law camp and
the power-law camp. Various attempts
\cite{princeton,93Bar,98SZ,WKG,RLP,Nagib} to validate the log-law
(\ref{loglaw}) or the alternative power-law~(\ref{BC}) were based on
extensive analysis of experimental data used to fit the velocity
profiles as a formal expansion in inverse powers of $\varepsilon$ or
as composite expansions in both $z^+$ and $\zeta$.

Recently a complementary approach to this issue was proposed on the
basis of experience with critical phenomena where one employs
scaling functions rather than scaling laws \cite{07LPR}. The essence of this
approach is the realization that a characteristic scale, say
$\~\ell$, may depend on the position in the flow. The simple scaling
assumption near the wall, $\~\ell ^+=\kappa z^+$, leads to the
log-law~\eq{loglaw}. The alternative suggestion of~\cite{93Bar},
$\~\ell^+ \propto (z^+)^{\alpha(\Ret)}$, leads to alternative
power-law~\eq{BC}. But there is no physical reason why $\~\ell$
should behave in either manner. Instead, it was shown that $\~\ell
/L$ should depend on $\zeta=z/L$, approaching $\kappa \zeta$ in the
limit $\zeta\to 0$ (in accordance to the classical thinking).
However for $\zeta\sim 1$, $\~\ell$ should saturate at some level
below $\kappa L $ due to the effect of other walls. We recall now
the recent analysis of DNS data that provides a strong support to
this idea, allowing one to get, within the traditional
(second-order) closure procedure, a quantitative description of the
following three quantities: the mean shear, $S(z)=d V(z)/dz$,   the
kinetic energy density (per unit mass), $K(z)\equiv  \langle |\B
u|^2\rangle/2$, and the tangential Reynolds stress, $W(z)\equiv  -\<
u_x u_z\>$. This is achieved in the entire flow and in a wide region
of $\Ret$, using only three $\Ret$-independent parameters.

The first relation between these objects follows from the
Navier-Stokes equation for the mean velocity. The resulting equation
is exact, being the mechanical balance between the momentum
generated at distance $z$ from the wall, i.e. $p'(L-z)$, and the
momentum transferred to the wall by kinematic viscosity and
turbulent transport. In physical and wall units it has the form:
\begin{equation}
\nu S + W= p'(L-z)\ \Rightarrow \ S^+ +W^+  =1-\zeta \ . \label{mom}
\end{equation}
Neglecting the turbulent diffusion of energy (known to be relatively
small in the log-low region), one gets a second relation as a local
balance between the turbulent energy generated by the mean flow at a
rate $SW$, and the dissipation at a rate $\varepsilon_{_K}\equiv \nu
\langle |\nabla u|^2\rangle$: $\varepsilon_{_K}\approx SW$. In
stationary conditions $\varepsilon_{_K}$ equals also the energy flux
from the outer scale of turbulence, $\~ \ell_{_K}$, toward smaller
scales. Thus flux is estimated as $\gamma_{_K}(z)K(z)$, where
$\gamma_{_K}(z)$ is the typical eddy turn over inverse time,
estimated as $\sqrt{K(z)}/ \~\ell_{_K} (z) $. This gives rise to the
other (now approximate) relations:
\begin{equation}
S^+ W^+\approx  \varepsilon_{_K}^+\,, \quad \varepsilon_{_K}^+=  \gamma _{_K}^+  \, K^+= K^+\, \sqrt{K^+}/\~\ell^+_{_K} \ . \label{ene}
\end{equation}
The third required relationship  can be obtained from the Navier
Stokes equation, similarly to \Eq{ene}, as the local balance between
the rate of Reynolds stress production $ \approx S K$ and its
dissipation $\varepsilon_{_W}$: $\varepsilon_{_W }\approx S K$. The
main contribution to $\varepsilon_{_W}$ comes from the so-called
Return-to-Isotropy process and can be estimated \cite{00Pope},
similarly to $\varepsilon_{_K}$, as $\gamma_{_W} \, W$ with
$\gamma_{_W}= \sqrt{K}/ \~\ell _{_W}$,
 involving yet another length-scale $\~\ell _{_W}$ which is of the same order of magnitude as $\ell_{_K}$.  Thus one has, similarly to \Eq{ene}:
 \begin{equation}
S^+ K^+\approx \varepsilon_{_W}^+\,, \quad \varepsilon_{_W}^+=
\gamma _{_W}^+ W^+ = W^+ \sqrt{K^+}/\~\ell^+_{_W} \ . \label{W}
 \end{equation}
 %%%%%%%%%%%%%%%%
\begin{figure*}
\includegraphics[width=0.3405 \textwidth]{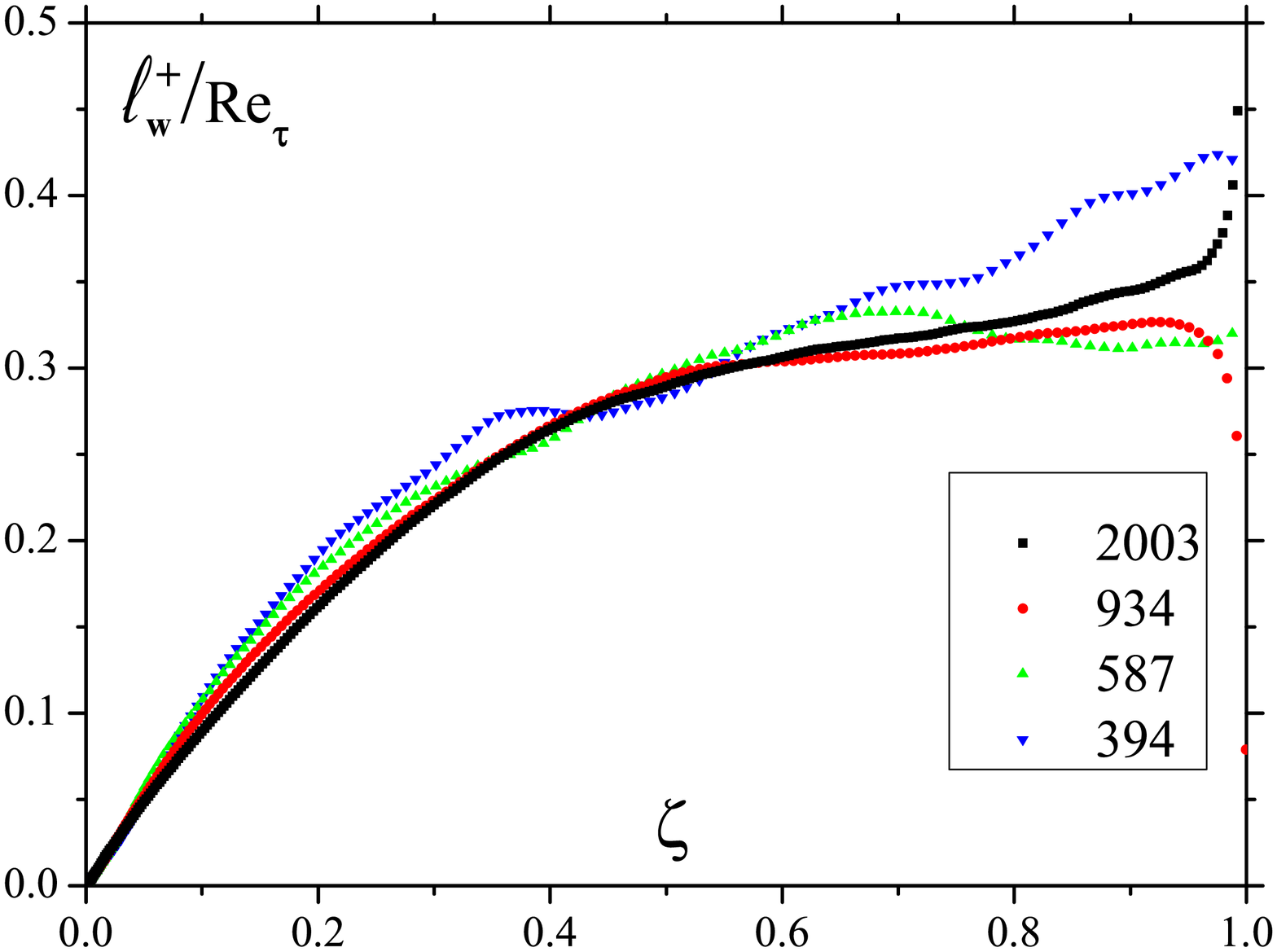}
\includegraphics[width=0.3405 \textwidth]{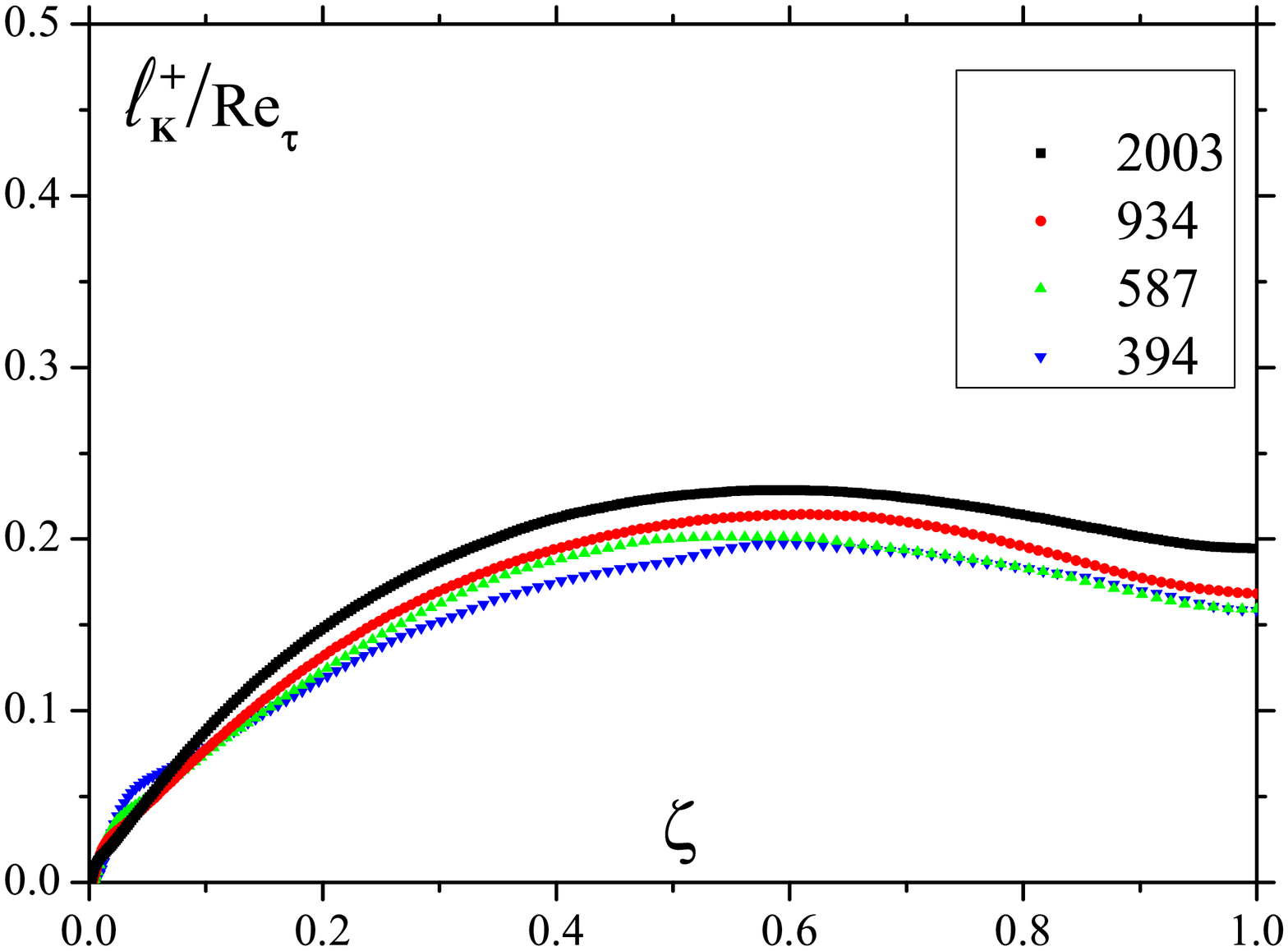}
\includegraphics[width=0.295 \textwidth]{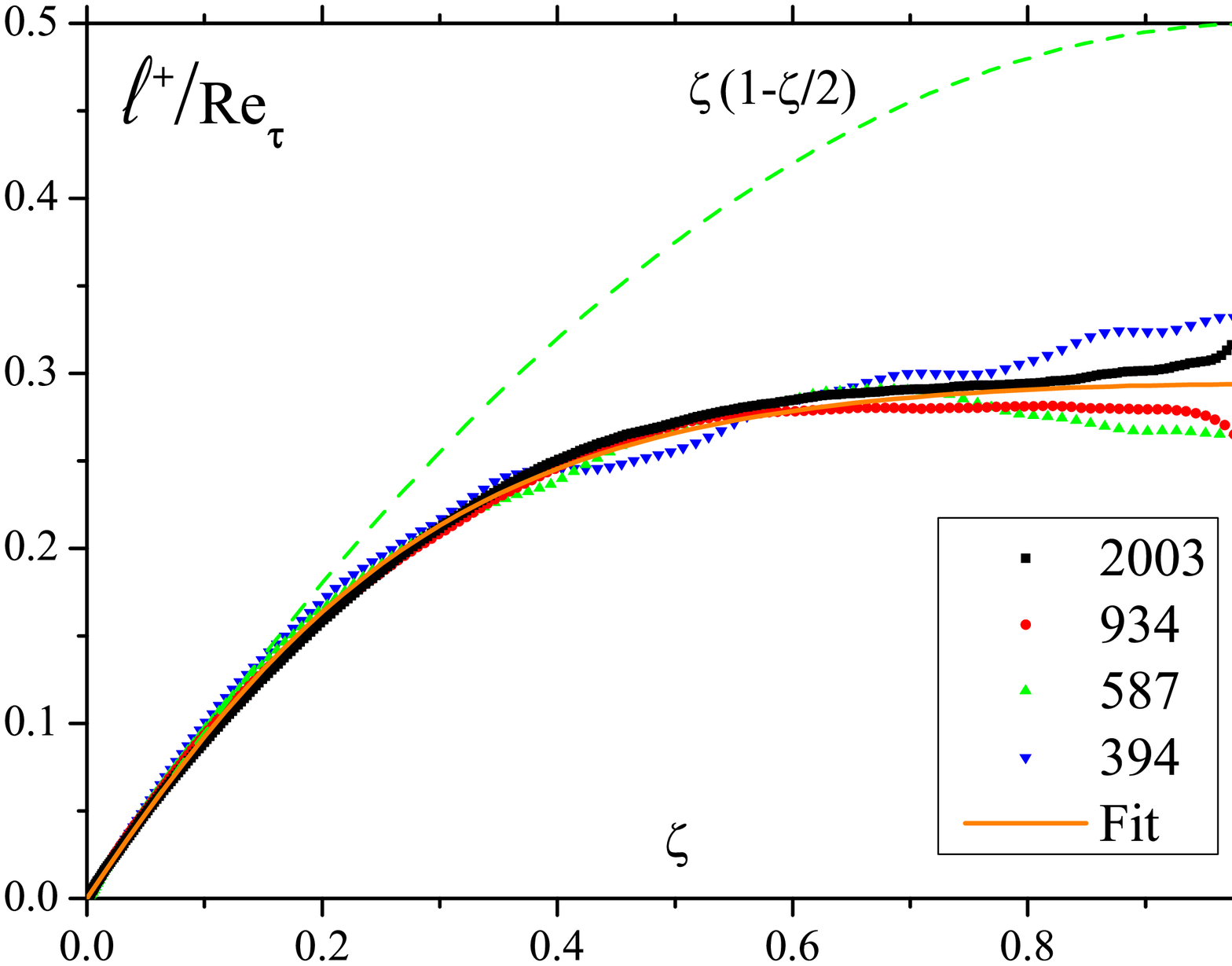}
\caption{Color online. The scaling function
$\ell_{_W}^+(\zeta)/\Ret$ (Left panel), $\ell^+_{_K}(\zeta)/\Ret$
(Middle panel) and the final scaling function $\ell^+(\zeta)$ (Right
panel), as a function of $\zeta\equiv z/L$, for four different
values of $\Ret$, computed from the DNS data \cite{Moser,DNS}. Note
the data collapse everywhere except at $\zeta\to 1$ where $W^+\sim
S^+ \ll 1$ and accuracy is lost. The green dash line represents
$\~\zeta=\zeta\,(1-\zeta/2)$ with a saturation level 0.5; in orange
solid line  we show the fitted function Eq.~(\ref{ell-fit}) with
$\ell\sb{sat}=0.311$.}
 \label{ell1}
\end{figure*}
%%%%%%%%%%%%%%%%%%%%%%%%%%%%%%%%%%%%%
Now  we show that the source of confusion is the assumption that the
length scales can be determined a priori as $\ell^+_{_{K,W}}\propto
(z^+)^\alpha$ with $\alpha=1$ or $\alpha\ne 1$. In reality we have
another characteristic length-scale, i.e. $L$, that also should
enter the game when $\zeta=z/L$ is not very small. The actual
dependence $\~\ell_{_W}$ and  $\~\ell_{_K}$ on $z$ and $L$ can  be
found from the data.  Consider first $\~\ell_{_W}$, defined by Eq.
(\ref{W}), and introduce a new scale $\ell_{_W}\equiv \~\ell_{_W}
r_{_W}(z^+)/\kappa_{_W} $ such that
\begin{equation}\label{defell1}
\ell_{_W}^+   \equiv  \frac{W^+(z^+,\Ret)\, r_{_W}(z^+)}{\kappa_{_W}
\, S^+(z^+,\Ret) \sqrt{K^+(z^+,\Ret)}}\ .
\end{equation}
Here, $r_{_W}(z^+)$ is a {universal, i.e. $\Ret$-independent
dimensionless {\it function} of $z^+$, chosen such that new scale
$\ell_{_W}/L=\ell_{_W}^+/ \Ret$ becomes a $\Ret$-independent
function of only one variable $\zeta  = ({z}/{L}) =({z^+}/{\Ret})$.
The dimensionless constant $\kappa_{_W}\approx 0.20$ is chosen to
ensure that $\lim _{z\ll L} \ell_{_W}^+(\zeta) =z^+$. Note that  if
$r_{_W}$ were a constant, $\ell_{_W}$ would have started near the
wall quadratically, i.e. as $z\times z^+$. Later  $\ell_{_W}^+$
would have become $\propto z^+$ for $50\ll z^+ \ll
\Ret$~\cite{00Pope}.  Thus to normalize it to slope 1 we need the
function $r_{_W}(z^+)$  that behaves $\propto 1/ z^+$ for $z^+\ll
50$ and approaches unity (under a proper choice of $\kappa_{_W}$)
for $z^+\gg 50$. A choice that leads to good data collapse reads
\begin{equation}\label{choice} r_{_W}(z^+)=\Big [1+\big( \ell\sb
{buf}^+\big / {z^+}\big)^6\Big ]^{1/6}\, , \quad
\ell\sb {buf}^+\approx 49 \,,
\end{equation}
where $\ell\sb {buf}^+ $ is a $\Ret$-independent length that plays a
role of the crossover scale (in wall units) between the buffer and
log-law region. The quality of the data collapse for this scaling
function is demonstrated in Fig.\ref{ell1}.

The second length-scale, $\~ \ell^+_{_K}$, is determined by
\Eq{ene}:%%
\begin{equation}\label{defell2}
\~\ell_{_K}^+ \equiv  \frac{(K^+(z^+,\Ret))^{3/2}} {
\varepsilon_{_K}^+(z^+,\Ret) }= \kappa_{_K}\ell_{_K}^+ \,, \quad
\kappa_{_K}\approx 3.7\ .
\end{equation}
In Fig.~\ref{ell1}  we demonstrate that this simple scaling function
leads to good data collapse everywhere except maybe in the viscous
layer. We will see below that this has only negligible effects on
our results.

 %%%%%%%%%%%%%%%%%%%%%%%%
 \noindent{\bf Solution and Velocity Profiles}:
Solving Eqs.\ (\ref{ene}), (\ref{W})  and accounting for
\Eqs{defell1}, \eq{defell2} we find
\begin{equation} W^+ = \big( \kappa \,S^+\ell^+
\big)^2 r_{_W}^{-3/2} \,, \label{solW} \end{equation} where we have
defined the von-K\`arm\`an constant and the crucial scaling function
$\ell^+(\zeta)$ as follows %%
\begin{equation}\label{defell} \kappa
\equiv ( \kappa_{_W}^3\kappa_{_K})^{1/4}\approx 0.415\,, \ \ell^+
\equiv [{\ell_{_W}^+}^3(\zeta)\, \ell^+_{_K}(\zeta)]^{1/4}  . %%
\end{equation}
The convincing data collapse for the resulting function
$\ell^+(\zeta)/\Ret$  is shown in Fig. \ref{ell1}, rightmost panel.
Substituting \Eq{solW} in \Eq{mom} we find a quadratic equation for
$S$ with a solution:
\begin{equation}%%
\label{S+}%%
S^+=\frac{\sqrt{1+{(1-\zeta)[2\kappa
\ell^+(\zeta)}]^2 \big/ r_{_W}(z^+)^{3/2}\,}-1}{2[\kappa\ell^+(\zeta)]^2
\big/ r_{_W}(z^+)^{3/2}} \ .
\end{equation}
To integrate this equation and find the mean velocity profile for
any value of $\Ret$ we need to determine the scaling function
$\ell^+(\zeta)$ from the data.  A careful analysis of the DNS data
allows us to find a good {\it one-parameter} fit for
$\ell^+(\zeta)$ \cite{footnote},
\begin{equation}
\label{ell-fit} \frac{\ell^+(\zeta)}{\Ret} = \ell\sb{s}\Big\{1
-\exp\Big[-\frac{\widetilde{\zeta}}{\ell_s}\,\Big(1
+\frac{\widetilde{\zeta}}{2 \ell_s} \Big)\Big ]\Big\}
\end{equation}
where $ \widetilde{\zeta}  \equiv  \zeta(1-\zeta/2)$ and $
\ell\sb{s} \approx 0.311 $. The quality of the fit is obvious from
the continuous line in the rightmost panel of Fig. \ref{ell1}.
 %%%%%%%%%%%%%%%%

%%%%%%%%%%%%%%%%%%%%

Finally the theory for the mean velocity contains three parameters,
namely $\ell\sb{s}$ together with  $\ell^+\sb{buff}$ and $\kappa$.
We demonstrate now that with these three parameters  we can
determine the mean velocity profile for any value $\Ret$, throughout
the channel, including the viscous layer, the buffer sub-layer, the
log-law region and the wake. Examples of the integration of Eq.
(\ref{S+}) are shown in Fig.~\ref{profiles}. We trust that
irrespective of the past adherence to the log-law camp or the
power-law camp, the sympathetic reader should agree that these fits
are very good. It remains now to estimate, using the explicit result
(\ref{S+}), when do we expect to see a log-law and when the
deviations due to a finite value of $\Ret$  would seem important. In
addition, our theory results also in the kinetic energy, and
Reynolds stress profiles which are in a qualitative agreement with
the DNS data; for $W$ profiles see Fig.~\ref{f:W}.

\begin{figure}
\includegraphics[width=0.45 \textwidth]{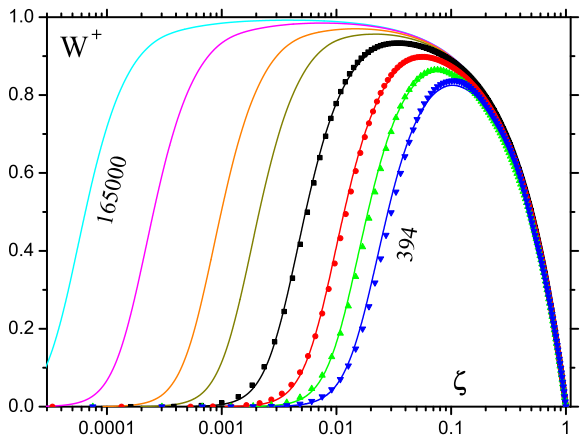}
\caption{\label{f:W} The Reynolds stress profiles (solid lines) at
$\Ret$ from 394 to 2003 (in channel) and from 5050 to 165,000 (in
pipe) in comparison with available DNS data (dots) for the channel.}
\end{figure}
%%%%%%%%%%%%%%%%%%%

To show that the present approach is quite general, we apply it now
to the experimental data that were at the center of the controversy
\cite{93Bar}, i.e., the Princeton Superpipe data \cite{princeton}.
In Fig.~\ref{profiles} right panel  we show the mean velocity
profiles as measured in the Superpipe compared with our prediction
using {\it the same scaling function} $\ell^+(\zeta)$. Note that the
data spans values of $\Ret$ from 5050 to 165000, and the fits with
only three $\Ret$-independent constants are excellent. Note the 2\%
difference in the value of $\kappa$ between the DNS and the
experimental data; we do not know at this point whether this stems
from inaccuracies in the DNS or the experimental data, or whether
turbulent flows in different geometries have different values of
$\kappa$. While the latter is theoretically questionable, we cannot
exclude this possibility until a better understanding of how to
compute $\kappa$ from first principles is achieved.

So far we discussed turbulent channel and pipe flows and
demonstrated the existence and usefulness of a scaling function
$\ell^+(\zeta)$ which allows us to get the profiles of the mean
velocities for all values of $\Ret$ and throughout the channel.
While this function begins near the wall as $z^+$, it saturates
later, and its full functional dependence on $\zeta$ is crucial for
finding the correct mean velocity profiles. The approach also allows
us to delineate the accuracy of the log-law presentation, which
depends on $z^+$ and the value of $\Ret$. For asymptotically large
$\Ret$ the region of the log-law can be very large, but nevertheless
it breaks down near the mid channel and near the buffer layer, where
correction to the log-law were presented.

The future challenge is to apply this idea to other examples
of wall-bounded flows, including time developing boundary layers,
turbulent flows with temperature gradients or laden with particles.
There may be more typical ``lengths" in such systems, and it is very
likely that turning these lengths into scaling functions will
provide new insights and better models for a variety of engineering
applications.

\section{Drag Reduction by Additives}
\label{drag}

One severe technological problem with turbulent flows is that they
cost a lot to maintain; the drag that the fluid exherts on the wall
increases significantly when turbulence sets in. It is therefore
important that there exist additives, in particular polymers and
bubbles, that can reduce this drag significantly. Over the last few
years there had been great progress in understanding these
phenomena, and here we provide a short review of this progress.

\subsection{Drag Reduction by Polymers}

The addition of few tens of parts per million (by weight) of
long-chain polymers to turbulent fluid flows in channels or pipes
can bring about a reduction of the friction drag by up to 80\%
\cite{49Toms,75Vir,97VSW,00SW}. This phenomenon of ``drag reduction"
is well documented and is used in technological applications from
fire engines (allowing a water jet to reach high floors) to oil
pipes. In spite of a large amount of experimental and simulations
data,  the fundamental mechanism for
drag reduction has remained under debate for a long time
\cite{69Lu,90Ge,00SW}. In such wall-bounded turbulence, the drag is
caused by momentum dissipation at the walls. For Newtonian flows (in
which the kinematic viscosity is constant) the momentum flux is
dominated by the so-called Reynolds stress, leading to the logarithmic
(von-K\'arm\'an) dependence of the mean velocity on the distance
from the wall \cite{00Pope}. However, with polymers, the drag
reduction entails a change in the von-K\'arm\'an log law such that a
much higher mean velocity is achieved. In particular, for high
concentrations of polymers, a regime of maximum drag reduction is
attained (the ``MDR asymptote"), independent of the chemical
identity of the polymer \cite{75Vir}, see Fig. \ref{profiles}.
%%%%%%%%%%%%%%%%%%%%%%%%%%%%%%%%%%%%%%%%%%%%%%%%%%%
\begin{figure}
\centering \epsfig{width=.40\textwidth,file=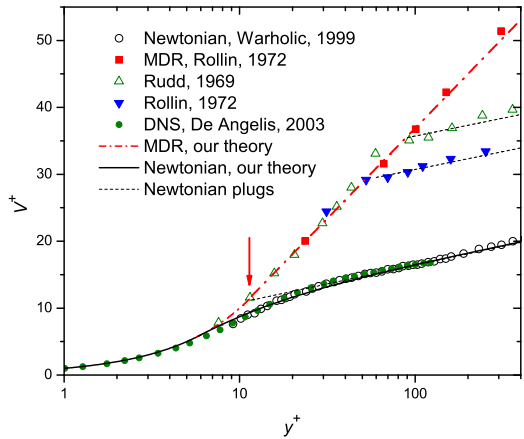}
\caption{Mean normalized  velocity profiles as a function of the
normalized  distance from the wall during drag reduction. The data
points from numerical simulations (green circles) \cite{03ACLPP} and
the experimental points (open circles) \cite{99WMH} represent the
Newtonian results. The black solid line is the universal Newtonian
line which for large $y^+$ agrees with von-K\'arm\'an's logarithmic
law of the wall (\ref{loglaw}). The red data points (squares)
\cite{72RS} represent the Maximum Drag Reduction (MDR) asymptote.
The dashed red curve represents our theory for the profile  which
for large $y^+$ agrees with the universal law (\ref{final}).  The
blue filled triangles \cite{72RS} and green open triangles
\cite{69Rud} represent the cross over, for intermediate
concentrations of the polymer,  from the MDR asymptote to the
Newtonian plug. Our theory is not detailed enough to capture this
cross over properly.} \label{profiles}
\end{figure}
%%%%%%%%%%%%%%%%%%%%%%%%%%%%%%%%%%%%%%%%%%%%%%%%%
During the last few years the fundamental mechanism for this
phenomenon was elucidated: while momentum is produced at a fixed
rate by the forcing, polymer stretching results in a suppression of
the momentum flux from the bulk to the wall. Accordingly the mean
velocity in the channel must increase. It was shown that polymer
stretching results in an effective viscosity that increases linearly
with the distance from the wall. The MDR asymptote is consistent
with the largest possible such linear increase in viscosity for
which turbulent solutions still exist. In other words, the MDR is an
edge solution separating turbulent from laminar flows. This insight
allowed one to derive the MDR as a new logarithmic law for the mean
velocity with a slope that fits existing numerical and experimental
data. The law is universal, explaining the MDR asymptote.

%%%%%%%%%%%%%%%%%%%%%%%%%%%%%%%%%%%%%%%%%%%%%%%%%%%%%%%%%%%%%%%%%%%%%%%%%%%%
\subsection{Short review of the theory }
The riddle of drag reduction can be introduced by a juxtaposition
of the effect of polymers with respect to the universal law
Newtonian law (\ref{loglaw}). In the presence of long chain polymers
the mean velocity profile $V^+(y^+)$ (for a fixed value of $p'$ and
channel geometry) changes dramatically. For sufficiently large
concentration of polymers $V^+(y^+)$ saturates to a new (universal,
polymer independent) ``law of the wall" \cite{75Vir},
\begin{equation}
V^+(y^+) = {\kappa_{_{\rm V}}}^{-1}\ln\left(e\, \kappa_{_{\rm V}}
y^+\right)\, \quad{\rm
for}~ y^+ \gtrsim 10
 \ . \label{final}
\end{equation}
This law, which was discovered experimentally by Virk (and hence the
notation $\kappa_{_{\rm V}}$), is also claimed to be universal,
independent of the Newtonian fluid and the nature of the polymer
additive, including flexible and rigid polymers \cite{97VSW}.
Previous to our work in this network, the numerical value of the
coefficient $\kappa_{_{\rm V}}$ was known only from experiments,
${\kappa_{_{\rm V}}}^{-1}\approx 11.7$, giving a phenomenological
MDR law in the form \cite{75Vir}
\begin{equation}
V^+(y^+) = 11.7\ln y^+ -17 \ . \label{finalexp}
\end{equation}

For smaller concentration of polymers the situation is as shown in
Fig. \ref{profiles}. The Newtonian law of the wall~(\ref{loglaw}) is
the black solid line for $y^+\gtrsim 30$. The MDR asymptote
(\ref{final}) is the dashed red line. For intermediate concentrations
the mean velocity profile starts along the asymptotic law
(\ref{final}), and then crosses over to the so called ``Newtonian
plug" with a Newtonian logarithmic slope identical to the inverse of
von-K\'arm\'an's constant. The region of values of $y^+$ in which
the asymptotic law~(\ref{final}) prevails was termed ``the elastic
sublayer" \cite{75Vir}. The relative increase of the mean velocity
(for a given $p'$) due to the existence of the new law of the wall
(\ref{final}) {\em is} the phenomenon of drag reduction. Thus the
main theoretical challenge is to understand the origin of the new
law~(\ref{final}), and in particular its universality, or
independence of the polymer used. A secondary challenge is to
understand the concentration dependent cross over back to the
Newtonian plug. In our work we argue that the phenomenon can be
understood mainly by the influence of the polymer stretching on the
$y^+$-dependent effective viscosity. The latter becomes a crucial
agent in carrying the momentum flux from the bulk of the channel to
the walls (where the momentum is dissipated by friction). In the
Newtonian case the viscosity has a negligible role in carrying the
momentum flux; this difference gives rise to the change of Eq.\
(\ref{loglaw}) in favor of Eq.\ (\ref{final}) which we derive below.

The equations of motion of polymer solutions are written in the
FENE-P approximation \cite{87BCAH,94BE} by coupling the fluid
velocity $\B u (\B r,t)$ to the tensor field of ``polymer
conformation tensor" $\B R (\B r,t)$. The latter is made from the
``end-to-end" separation vector as $ R_{\alpha\beta}(\B r,t)\equiv
\langle r_\alpha r_\beta \rangle$, and it satisfies the equation of
motion
\begin{eqnarray}
\frac{\partial R_{\alpha\beta}}{\partial t}+( u_{\gamma}  \nabla_{\gamma})
R_{\alpha\beta}
&&=\frac{\partial u_\alpha}{\partial r_\gamma}R_{\gamma\beta}
+R_{\alpha\gamma}\frac{\partial u_\beta}{\partial r_\gamma}\nonumber\\
&&-\frac{1}{\tau}\left[ P(\B r,t) R_{\alpha\beta} -\rho_0^2
\delta_{\alpha\beta} \right]\ ,\nonumber
\end{eqnarray}
\begin{equation}
P(\B r,t)=(\rho_m^2-\rho_0^2)/(\rho_m^2 -R_{\gamma\gamma})
\end{equation}
$\rho^2_m$ and $\rho^2_0$ refer to the maximal and the equilibrium
values of the trace $R_{\gamma\gamma}$. In most applications
$\rho_m\gg \rho_0$ $$P( r,t)\approx (1/(1 -\alpha
R_{\gamma\gamma})$$ where $\alpha=\rho_m^{-2}$. The equation for the
fluid velocity field gains a new stress tensor:
\begin{equation}
\frac{\partial  u_{\alpha}}{\partial t}+( u_{\gamma}\nabla_{\gamma})
u_{\alpha}=-\nabla_{\alpha} p +\nu_s \nabla^2  u_{\alpha} +
\nabla_{\gamma} {T_{\alpha\gamma}}
\end{equation}
\begin{equation}{T_{\alpha\beta}}( r,t) =
\frac{\nu_p}{\tau}\left[\frac{P( r,t)}{\rho_0^2}
R_{\alpha\beta}(r,t) - \delta_{\alpha\beta} \right] \ .
\end{equation}
Here $\nu_s$ is the viscosity of the neat fluid, and $\nu_p$ is a
viscosity parameter which is related to the concentration of the
polymer, i.e. $\nu_p/\nu_s\sim c_p$.\\~\\ We shall use the
approximation
$$
    T_{\alpha\beta} \sim \frac{\nu_p}{\tau}\frac{P}{\rho_0^2} R_{\alpha\beta}.
$$

Armed with the equation for the viscoelastic medium we establish the
mechanism of drag reduction following the standard strategy of
Reynolds. Eq.\ (\ref{mom}) changes now to another exact relation
\cite{04BDLPT} between the objects  $S$ and $W$ which includes the
effect of the polymers:
\begin{equation}
\label{MF}
W + \nu S + \frac{\nu_p}{\tau} \langle P R_{xy}\rangle (y) =
p'(L-y) \ .
\end{equation}
On the RHS of this equation we see the production of momentum flux
due to the pressure gradient; on the LHS we have the Reynolds
stress, the Newtonian viscous contribution to the momentum flux, and
the polymer contribution to the momentum flux. A second relation
between $S(y)$, $W(y)$, $K(y)$ and $\B R(y)$ is obtained from the
energy balance. In Newtonian fluids the energy is created by the
large scale motions at a rate of $W(y) S(y)$. It is cascaded down
the scales by a flux of energy, and is finally dissipated at a rate
$\epsilon$, where $\epsilon = \nu \langle |\nabla u|^2\rangle$. In
viscoelastic flows one has an additional contribution due to the
polymers. Our calculation \cite{04BDLPT} showed that the energy
balance equation takes the form:
\begin{equation}
\label{EB}
a\nu\frac{K}{y^2}+ b\frac{K^{3/2}}{y}+\frac{A^2 \nu_p}{2
\tau^2} \langle P\rangle^2(\langle R_{yy}\rangle +\langle
R_{zz}\rangle) = WS \ .
\end{equation}
We note
that contrary to Eq. (\ref{MF}) which is exact, Eq.(\ref{EB}) is
not exact. We expect it however to yield good order of magnitude
estimates as is demonstrated below.  Finally, we quote the
experimental fact \cite{75Vir} that outside the viscous
boundary layer
\begin{equation}
\label{WK} \frac{W(y)}{K(y)} = \Bigg \{ \begin{array}{ll} c\Sub
N^2\,,& \text {for Newtonian flow,}\\ & \\
c\Sub V^2\,, & \text {for viscoelastic flow.} \\
\end {array} \end{equation}%%
The coefficients $c\Sub{N}$ and $c\Sub{V}$ are bounded from above by
unity. (The proof is $|c|\equiv |W|/K\le 2|\langle u_xu_y\rangle|/
\langle u_x^2+u_y^2\rangle\le 1$, because $(u_x\pm u_y)^2\ge 0$.)

To proceed, one needs to estimate the various components of the
polymer conformation tensor. This was done in \cite{04LPPTa} with
the final result that for $c_p$ large (where  $P\approx 1$), and
Deborah number ${\rm De}\equiv \tau S(y)\gg 1$ the conformation
tensor is highly anisotropic,
\begin{equation}
    \B R(y) \simeq R^{yy}(y)\left(\begin{array}{ccc}
        2{\rm De} ^ 2 (y)& ~{\rm De}(y)~ &  ~~0~~\\
       {\rm De}(y) & 1 & 0 \\
        0 & 0 & C(y)
    \end{array}\right)\ .\nonumber
    \end{equation}

The important conclusion is that the term proportional to $\langle
R_{yy}\rangle$ in Eq.\ (\ref{EB}) can be written as $\nu_p\langle
{\cal R}_{yy}\rangle(y) S(y)$. Defining the ``effective viscosity"
$\nu(y)$ according to
\begin{equation}
\nu(y) = \nu_0 + \nu_p \langle {\cal R}_{yy} \rangle(y) \ ,
\end{equation}
The momentum balance equation attains the form
\begin{equation}
\nu(y) S(y) + W(y) = p'L \ . \label{vismom}
\end{equation}
It was shown in \cite{04BDLPT} that also the energy balance equation
can be rewritten with the very same effective viscosity, i.e.,
\begin{equation}
\nu (y)\Big( \frac{a}{y} \Big)^2 K(y) +\frac{b\,\sqrt{K(y)}}y
K(y) = W(y)S(y)\ . \label{visen}
\end{equation}
In the MDR region the first term on the RHS of in Eqs.\
(\ref{vismom}) and (\ref{visen}) dominate; from the first equation
$\nu(y)\sim 1/S(y)$. Put in Eq.\ (\ref{visen}) this leads to
$S(y)\sim 1/y$, which translates to the new logarithmic law which is
the MDR. We will determine the actual slope momentarily. At this
point one needs to stress that this results means that $nu(y)$ must
be proportional to $y$ in the MDR regime. This linear dependence of
the effective viscosity is one of the central discoveries of our
approach. Translated back, it predicts that $\langle R_{yy} \rangle
\sim y$ outside the boundary layer. This prediction is well
supported by numerical simulations.

The crucial new insight that explained the universality of the MDR
and furnished the basis for its calculation is that the MDR  is a
marginal flow state of wall-bounded turbulence: attempting to
increase $S(y))$ beyond the MDR results in the collapse of the
turbulent solutions in favor of a stable laminar solution $W=0$. As
such, the MDR is universal by definition, and the only question is
whether a polymeric (or other additive) can supply the particular
effective viscosity $\nu(y)$ that drives Eqs.\ (\ref{vismom}) and
(\ref{visen}) to attain the marginal solution that maximizes the
velocity profile. We predict that the same marginal state will exist
in numerical solutions of the Navier-Stokes equations furnished with
a $y$-dependent viscosity $\nu(y)$. There will be no turbulent
solutions with velocity profiles higher than the MDR.

To see this explicitly, we first rewrite the balance equations in
wall units. For constant viscosity (i.e. $\nu(y) \equiv \nu_0$),
Eqs.\ (\ref{vismom})-(\ref{visen}) form a closed set of equations
for $S^+ \equiv S \nu_0/(P'L)$ and $W^+ \equiv W/\sqrt{P'L}$ in
terms of two dimensionless constant ${\delta^+} \equiv a\sqrt{K/W}$
(the thickness of the viscous boundary layer) and $\kappa_{_{\rm K}}
\equiv b/c_V^3$ (the von K\'arm\'an constant). Newtonian experiments
and simulations agree well with a fit using $\delta^+ \sim 6$ and
$\kappa_{_{\rm K}}  \sim 0.436$ (see the black continuos line in
Fig.\ \ref{profiles} which shows the mean velocity profile using
these very constants). Once the effective viscosity $\nu(y)$ is no
longer constant we expect $c_V$ to change and consequently the two
dimensionless constants will change as well. We will denote the new
constants as $\Delta$ and $\kappa_{_{\rm C}} $ respectively. Clearly
one must require that for $\nu(y)/\nu_0 \rightarrow 1$, $\Delta
\rightarrow \delta^+$ and $\kappa_{_{\rm C}}  \rightarrow
\kappa_{_{\rm K}} $. The balance equations are now written as
\begin{eqnarray}
&&\nu^+(y^+) S^+ (y^+) +W^+(y^+)  =1\ , \label{bal1}\\
&&\nu^+(y^+)\frac{{\Delta}^2}{{y^+}^2
}+\frac{\sqrt{W^+}}{\kappa_{_{\rm C}}y^+} =S^+\  . \label{bal2}
\end{eqnarray}
where  $\nu^+(y^+)\equiv \nu(y^+)/\nu_0$. Substituting now $S^+$
from Eq.\ (\ref{bal1}) into Eq.\ (\ref{bal2}) leads to a quadratic
equation for $\sqrt{W^+}$. This equation has as a zero solution for
$W^+$ (laminar solution) as long as $\nu^+(y^+){\Delta}/y^+=1$.
Turbulent solutions are possible only when $\nu^+(y^+){\Delta}/y^+
<1$. Thus at the edge  of existence of turbulent solutions we find
$\nu^+\propto y^+$ for $y^+ \gg 1$. This is not surprising, since it
was observed already in previous work that the MDR solution is
consistent with an effective viscosity which is asymptotically
linear in $y^+$ \cite{04LPPT,04DCLPPT}. It is therefore sufficient
to seek the edge solution of the velocity profile with respect to
linear viscosity profiles, and we rewrite Eqs.\ (\ref{bal1}) and
(\ref{bal2}) with an effective viscosity that depends linearly on
$y^+$ outside the boundary layer of thickness $\delta^+$:
\begin{eqnarray}
&&[1+\alpha(y^+-\delta^+)]S^+ +W^+ =1\ ,\label{pr1}\\
&&[1+\alpha(y^+-\delta^+)]\frac{{\Delta}^2(\alpha)}{{y^+}^2
}+\frac{\sqrt{W^+}}{\kappa_{_{\rm C}}  y^+} =S^+\  . \label{pr2}
\end{eqnarray}

We now endow $\Delta$ with an explicit dependence on the slope of
the effective viscosity  $\nu^+(y)$, $\Delta = \Delta(\alpha)$.
Since drag reduction must involve a decrease in $W$, we expect the
ratio $a^2 K/W$ to depend on $\alpha$, with the constraint that
$\Delta(\alpha)\to \delta^+$ when $\alpha\to 0$. Although $\Delta$,
$\delta^+$ and $\alpha$ are all dimensionless quantities, physically
$\Delta$ and $\delta^+$ represent (viscous) length scales (for the
linear viscosity profile and for the Newtonian case respectively)
while $\alpha^{-1}$ is the scale associated to the slope of the
linear viscosity profile. It follows that $\alpha \delta^+$ is
dimensionless even in the original physical units. It is thus
natural to present $\Delta(\alpha)$ in terms of a dimensionless
scaling function $f(x)$,
\begin{equation}
\Delta(\alpha) =\delta^+ f(\alpha\delta^+) \ . \label{scaling}
\end{equation}
Obviously, $f(0)=1$. In \cite{05BALP} it was shown that the balance
equation (\ref{pr1}) and (\ref{pr2}) (with the prescribed form of
the effective viscosity profile) have an non-trivial symmetry  that
leaves them invariant under rescaling of the wall units. This
symmetry dictates the function $\Delta(\alpha)$ in the form
\begin{equation}
\Delta(\alpha) =\frac{\delta^+}{1-\alpha\delta^+} \  . \label{Delta}
\end{equation}
Armed with this knowledge we can now find the maximal possible
velocity far away from the wall, $y^+\gg \delta^+$. There the
balance equations simplify to
\begin{eqnarray}
&&\alpha y^+S^+ +W^+ =1\ ,\label{pr12}\\
&&\alpha \Delta^2(\alpha)+
\sqrt{W^+}/\kappa_{_{\rm C}} =y^+S^+\  . \label{pr22}
\end{eqnarray}
These equations have the $y^+$-independent solution for $\sqrt{W^+}$ and
$y^+S^+$:
\begin{eqnarray}
\sqrt{W^+} &=&-\frac{\alpha}{2\kappa_{_{\rm C}}} +
\sqrt{\Big(\frac{\alpha}{2\kappa_{_{\rm C}}}\Big)^2
+1-\alpha^2\Delta^2(\alpha)} \ , \nonumber\\
y^+S^+&=&\alpha \Delta^2(\alpha)+\sqrt{W^+}/\kappa_{_{\rm C}} \ . \label{quadsol}
\end{eqnarray}
By using equation (\ref{quadsol}) (see Fig. \ref{Fig1}), we obtain
that the edge solution ($W^+ \rightarrow 0$) corresponds to the
supremum of $y^+S^+$, which happens precisely when $\alpha=1/
\Delta(\alpha)$. Using Eq.\ (\ref{Delta}) we find the solution
$\alpha=\alpha_m =1/2\delta^+$. Then $y^+S^+=\Delta(\alpha_m)$,
giving $\kappa^{-1}_{_{\rm V}}=2\delta^+$. Using the estimate
$\delta^+\approx 6$ we get the final prediction for the MDR. Using
Eq.\ (\ref{final}) with $\kappa^{-1}_{_{\rm V}}=12$, we get
%%%%%%% FIGURE 2 %%%%%%%%%%%%%%%%%%
\begin{figure}
\centering \epsfig{width=.50\textwidth,file=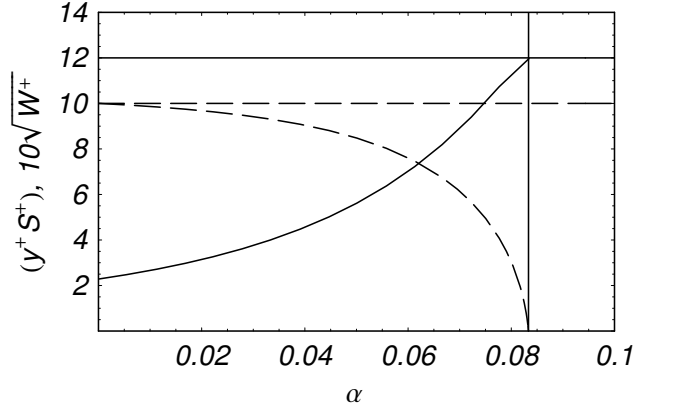}
\caption{The solution for 10$\sqrt{W^+}$ (dashed line) and $y^+S^+$
(solid line) in the asymptotic region $y^+\gg \delta^+$, as a
function of $\alpha$. The vertical solid line
$\alpha=1/2\delta^+=1/12$ which is the edge of turbulent solutions;
Since $\sqrt{W^+} $ changes sign here, to the right of this line
there are only laminar states. The horizontal solid line indicates
the highest attainable value of the slope of the MDR logarithmic law
$1/\kappa_{_{\rm V}}=12.$} \label{Fig1}
\end{figure}
%%%%%%%%%%%%%%%%%%%%%%%%%%%%%%%%%%
\begin{equation}
V^+(y^+) \approx 12\ln{y^+}  -17.8 \ . \label{predicta}
\end{equation}

This result is in close agreement with the empirical law
(\ref{finalexp}) proposed by Virk. The value of the intercept on the
RHS of Eq.\ (\ref{predicta}) follows from Eq.\ (\ref{final}) which
is based on matching the viscous solution to the MDR log-law in
\cite{04LPPT}. Note that the numbers appearing in Virk's law
correspond to $\delta^+= 5.85$, which is well within the error bar
on the value of this Newtonian parameter. Note that we can easily
predict where the asymptotic law turns into the viscous layer upon
the approach to the wall. We can consider an infinitesimal $W^+$ and
solve Eqs.\ (\ref{bal1}) and (\ref{bal2}) for $S^+$ and the
viscosity profile. The result, as before, is
$\nu^+(y)=\Delta(\alpha_m) y^+$. Since the effective viscosity
cannot fall bellow the Newtonian limit $\nu^+=1$ we see that the MDR
cannot go below $y^+=\Delta(\alpha_m)=2\delta^+$. We thus expect an
extension of the viscous layer by a factor of 2, in very good
agreement with the experimental data.

%%%%%%%%%%%%%%%%%%%%%%%%
\subsection{Non-universal aspects of drag reduction by polymers}

When the concentration of polymers is not large enough, or when the
Reynolds number is too low, the MDR is attained only up to some
value of $y^+$ that depends in a non-universal manner on the
Reynolds number and on the nature of the polymer \cite{02CLLC}. These
non-universal turn-backs to the so called ``Newtonian plug" can be
understood theoretically, and we refer the reader to \cite{04BLPT,06BALPT} for
further details.

\subsection{Drag reduction by micro-bubbles}

Finally, we should mention that drag reduction by polymers is not
the solution for many technologically pressing problems, the most
prominent of which is the locomotion of ships. Here a more promising
possibility is the drag reduction by bubbles, a subject that is much
less developed than drag reduction by polymers. For some recent
papers on this subject  see for example  \cite{06LLP} and references therein; 
we we stress that this subject is far from being exhausted by these papers, and we expect that more work should appear on this subject in the near future.

%%%%%%%%%%%%%%%%%%%%%%%%%%%%%%%%%%%
\section{Thermal Convection}
\label{convection}

Convection often occurs in conjunction with other features such as
rotation, magnetic field and particulate matter, so the knowledge of
the subject is relevant to several closely related fields. The
complexity of the underlying equations has precluded much analytical
progress, and the demands of computing power are such that routine
simulations of large turbulent flows has not yet been possible.
Thus, the progress in the field has depended more on input from
experiment, which itself has severe limitations. The progress in the
subject, such as it is, has been possible only through strong
interactions among theory, experiment and simulation. This is as it
should be.

The paradigm for thermal convection is the Rayleigh-B\'enard problem
in which a thin fluid layer of infinite lateral extent is contained
between two isothermal surfaces with the bottom surface maintained
slightly hotter. When the expansion coefficient is positive (as is
the case usually), an instability develops because the hot fluid
from below rises to the top and the colder fluid from above sinks to
the bottom. The applied driving force is measured in terms of a
Rayleigh number, $Ra$,
\begin{equation}
Ra = g \alpha \Delta T H^3/ \nu \kappa,
\end{equation}
which emerges in front of the buoyancy term and is a non-dimensional
measure of the imposed temperature difference across the fluid
layer. Here, $g$ is the acceleration due to gravity, $H$ is the
vertical distance between the top and bottom plates, $\alpha$, $\nu$
and $\kappa$ are, respectively, the isobaric thermal expansion
coefficient, the kinematic viscosity and the thermal diffusivity of
the fluid. Physically, the Rayleigh number measures the ratio of the
rate of potential energy release due to buoyancy to the rate of its
dissipation due to thermal and viscous diffusion.

The second important parameter is the Prandtl number
\begin{equation}
Pr = \nu/\kappa,
\end{equation}
which is the ratio of time scales due to thermal diffusion
($\tau_\theta = H^2/\kappa$) and momentum diffusion ($\tau_v=
H^2/\nu$), and determines the ratio of viscous and thermal boundary
layers on the solid surfaces. With increasing $Ra$ the dynamical
state of the Rayleigh-B\'enard system goes from a uniform and
parallel roll pattern at the onset ($Ra \sim 10^3$) to turbulent
state at $Ra \sim 10^7 - 10^8$. (The onset value is independent of
$Pr$ but the latter depends strongly on it.)

For purposes of theoretical simplification, it is customary to
assume that the thermal driving does not affect the pressure or the
incompressibility condition, and that its only effect is to
introduce buoyancy. This is the Boussinesq approximation. How
closely the theoretical results correspond to observations depends
on how closely the experiments obey the Boussinesq approximation. It
is also not clear if small deviations from the ideal boundary
conditions produce only small effects.

\subsection{Experiments using cryogenic helium}

Since many examples of convection occur at very high Rayleigh
numbers \cite{sree-russ}, it is of interest to understand the heat
transport characteristics in that limit. It is also necessary to be
able to cover a large range of $Ra$ in order to be able to discover
the applicable scaling laws. Cryogenic helium has been used
successfully for the purpose. Though experiments in conventional
fluids have been valuable \cite{gold,ahle}, the Rayleigh number has
been pushed to the limit only through cryogenic helium. The same
properties that make it a suitable fluid for convection studies also
makes it suitable for creating flows with very high Reynolds numbers
\cite{smit}.

Historically, a small ``superfluid wind tunnel" was constructed
\cite{crai} with the idea of exploiting the superfluid properties of
helium II for obtaining very high Reynolds numbers. Potential flow
was observed for low velocities, with no measurable lift on a pair
of fly wings hanging in the tunnel, but the inevitable appearance of
quantized vortices (see section VII on superfluid turbulence)
altered that picture for higher flow speeds. Threlfall \cite{thre}
recognized the advantages of using low temperature helium gas to
study high-$Ra$ convection. The later work by Libchaber and
co-workers \cite{cast} brought a broader awareness of the potential
of helium. The work of Refs.\ \cite{natu,chav} is a natural
culmination of this cumulative effort.

\begin{center}
\begin{tabular}{|c|c|c|c|}
\hline $Fluid$&$T (K)$&$P (Bar)$&$\alpha/\nu \kappa$ \\
\hline
Air&293&1&0.12\\
Water&293&1&14\\
Helium I&2.2&SVP&$2.3\times 10^{5}$\\
Helium II&1.8&SVP&$---$\\
Helium gas&5.25&2.36&$6\times 10^{9}$\\
Helium gas&4.4&$2\times 10^{-4}$&$6\times 10^{-3}$\\
\hline
\end{tabular}
\\
\vspace{5mm} Table 1. Values of the combination of fluid properties
$\alpha/\nu \kappa$ for air, water and helium, From Ref.\
\cite{jltp}. \vspace{3mm}
\end{center}

The specific advantage of using helium for convection is the huge
value of the combination $\alpha/\nu \kappa$ near the critical
point. This can generate large $Ra$ (see Table 1). For a fluid layer
some 10 meters tall and a reasonable temperature difference of
$0.5K$, Rayleigh numbers of the order $10^{21}$ are possible. Table
1 also shows that $\alpha/\nu \kappa$ is quite small at pressures
and temperatures sufficiently far away from the critical value. In
fact, the range shown in the table covers a factor of $10^{12}$, so
any experiment of fixed size $H$ can yield at least 12 decades of
the control parameter $Ra$ by this means alone. However, if $H$ is
chosen to be large enough, this entire range of $Ra$ can be shifted
to a regime of developed turbulence where well-articulated scaling
relations may be observed. This tunability is essentially impossible
for air and water, especially because one cannot use more than
modest temperature difference to increase $Ra$ (due to the attendant
non-Boussinesq effects, Sect.\ VI C). For other advantages in using
helium, see \cite{jltp}.

\subsection{The scaling of the heat transport}

The heat transport in convection is usually given in terms of the
Nusselt number $Nu$
\begin{equation}
Nu = \frac{q}{q_{cond}} = \frac{qH}{k_{f} \Delta T},
\end{equation}
where $q$ is the total heat flux, $q_{cond}$ is the heat flux in the
absence of convection, given by Fourier's law, and $k_{f}$ is the
thermal conductivity of the fluid. $Nu$ represents the ratio of the
effective turbulent thermal conductivity of the fluid to its
molecular value. One goal of convection research is to determine the
functional relation $Nu = f(Ra,Pr)$. This relation is at least as
fundamental as the skin friction relation in isothermal flows.

Figure 6, reproduced from Ref.\ \cite{jltp}, illustrates the
enormous range of $Ra$ and $Nu$ that is possible in low temperature
experiments of modest physical size. The Nusselt numbers have been
corrected here for sidewall conduction and also for finite thermal
conductivity of the plates (and both corrections are small, see
\cite{jltp}). That one can reach Nusselt numbers as high as $10^4$
bears testimony to the great importance of turbulence as a subject
of serious study.

We have shown this figure in part because it represents the highest
$Ra$ achieved so far under laboratory conditions and also the
largest range of $Ra$ in the turbulent scaling regime, both of which
represent the fulfilment of the promise of cryogenic helium gas. The
average slope over 11 decades is 0.32, close to 1/3. In part, we
show the graph because one might have hoped that such an unusual
figure spanning many decades in $Ra$ might have a finality to it.
Perhaps it does. However, experiments of Chavanne et al.\
\cite{chav}, and by Niemela \& Sreenivasan \cite{NandS_JFM} for a
different aspect ratio, have found a scaling exponent rising beyond
1/3 towards the very highest $Ra$. The plausible conclusions of
Niemela \& Sreenivasan \cite{NandS_JFM,ns} were that those data
corresponded to large departures from Boussinesq conditions and to
variable Prandtl number (remembering that increasing Prandtl number
serves to stabilize and laminarize the boundary layers, in contrast
to conditions prescribed for observing the 1/2-power scaling), but
it is important to test these plausible conclusions directly. We
shall momentarily discuss the current work in this direction. If we
ignore the apparently non-Boussinesq regime, it has been argued in
Refs.\ \cite{NandS_JFM,ns} that the scaling exponent from existing
data is most likely consistent with a value close to 1/3.

As already mentioned, computations have not yet approached
experiments in terms of high $Ra$, but their advantage is that $Pr$
can be held constant and the Boussinesq approximation can be
enforced strictly. The limit of computational ability has recently
been pushed by Amati et al.\ \cite{amat}, who have reached Rayleigh
numbers of $2\times 10^{14}$. Even though this number is still about
three orders of magnitude lower than the highest experimental value,
it has become quite competitive with respect to many other
experiments. This work suggests that the one-third exponent is quite
likely, reinforcing the conclusion of Refs.\ \cite{NandS_JFM,ns}.
Computational simulations have also explored the effects of finite
conductivity, sidewall conduction and non-Boussinesq effects
\cite{verz,same}.

In spite of the limitations of $Ra$ attainable in simulations, much
of the detail we know about boundary layers and fluctuations come
from them. Direct knowledge of the velocity is most desirable in
understanding the dynamics of plumes and boundary layers, and also
the importance of the mean wind. Experiments in convection have
limited themselves to measuring the mean wind and temperature at a
few points, but not the spatial structures. The conventional
techniques of velocity measurements and flow visualization are
fraught with difficulties, as has been discussed in \cite{jltp}.

We should now discuss the contributions of the theory to the heat
transport problem. Two limiting cases for the scaling of $Nu$ have
been considered. The first scenario imagines that the global flux of
heat is determined by processes occurring in the two thermal
boundary layers at the top and bottom of the heated fluid layer.
Then the intervening turbulent fluid, being fully turbulent and
``randomized", acts as a thermal short circuit and therefore its
precise nature is immaterial to the heat flux. We can then determine
the relation to be $Nu \sim Ra^{1/3}$ \cite{malk54}. This scaling
assumes that the heat flux has no dependence on $H$. In the limit in
which molecular properties are deemed irrelevant in determining heat
transport---that is, when boundary layers cease to exist---an
exponent of 1/2 (modulo logarithmic corrections) has been worked out
phenomenologically \cite{krai2}. There has been an alternative
theory \cite{cast} that obtains the $2/7$-ths scaling through
intermediate asymptotics, but the experimental result that motivated
the work has not been sustained by more recent work.

The upperbound theory, though quite old (see Refs.\
\cite{malk54,howa}), has been taken to new levels through the
efforts of Constantin and Doering (e.g., Ref.\ \cite{cons}), as well
as by others more recently, and has contributed some valuable hints
on the heat transport law. The latest summary is as follows:

1. Arbitrary Prandtl number: $Nu < Ra^{1/2}$ uniformly in Prandtl
number \cite{cons}. This result rules out the Prandtl number
dependence such as $Pr^{1/2}$ \cite{spie,GL} and $Pr^{-1/4}$
\cite{krai2}. In particular, the latter paper was written when the
boundary layer structure was understood much less, and there is a
need to reconstruct its arguments afresh, in particular for the
reassessment of the Rayleigh number at which the so-called
``ultimate regime" is supposed to prevail for Prandtl numbers of
order unity.

2. Large but finite Prandtl number: The largeness of the Prandtl
number is prescribed by the condition $Pr > c Ra$, where $c$ is a
constant of the order unity. Under this condition, the upperbound is
given by $Nu \le Ra^{1/3} (ln Ra)^{2/3}$ \cite{wang}. For higher
Rayleigh numbers the upperbound is still given by (1) above.

3. Infinite Prandtl number: The latest result due to Doering et al.\
\cite{doer}, is $Nu \le CRa^{1/3}(ln Ra)^{1/3}$. Robust calculations
by Ireley et al.\ \cite{irel}, which still seem to fall short of
proof, is $Nu \le a Ra^{1/3}$, where $a$ is a constant of the order
unity.

Thus, as far as the upperbound theory goes, the $Ra^{1/2}$ result is
permissible for Prandtl numbers of the order unity, though some
semi-analytical results on Prandtl number dependencies are ruled out
as noted above.

Finally, we mention the effect of rough surfaces on the global heat
transfer rate \cite{tong,roch} and the presence of a weakly
organized mean wind \cite{wind,sbn,qui01,vill}. These studies have
added to our understanding of turbulent convection. The wind
phenomenon has had a rather broad reach; e.g., quantitative
observations of occasional reversals of the mean wind flow direction
have been shown to be related to simple models of self-organized
criticality \cite{soc}. Furthermore, the lifetimes of the metastable
states of the bi-directional mean flow have intriguing analogies
with reversals of the Earth's magnetic field polarity, a phenomena
arising from turbulent convection within the outer core \cite{glat};
there is also a quantitative statistical analogy with the lifetime
of solar flare activity driven by turbulent convection in the Sun's
outer layer \cite{sola}. This latter conclusion may indicate the
existence of an underlying universality class, or a more direct
physical similarity in the convective processes that lead to
reshuffling of the magnetic footprints and ultimately to flare
extinction.

%%%%%%%%%%%%%%%%%%%%%%%%%%%%%%%
\begin{figure}
\begin{center}
\includegraphics[width=.50\textwidth]{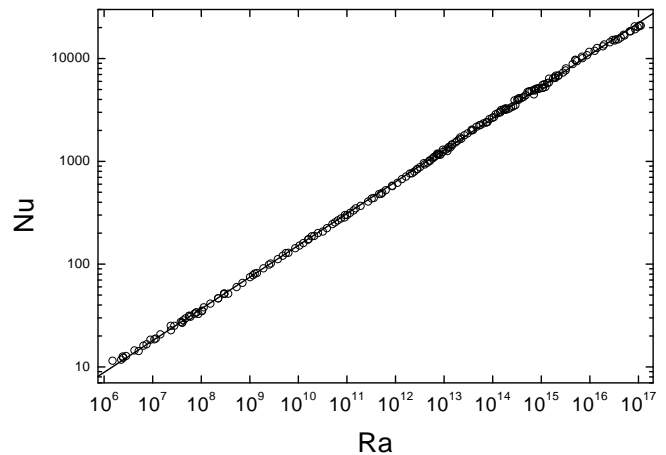}
\caption{Log-log plot of the Nusselt number versus Rayleigh number.
The line through the data is a least-square fit over the entire Ra
range, and represents a $dlogNu/dlogRa$ slope of 0.32.}
\label{figure6}
\end{center}
\end{figure}
%%%%%%%%%%%%%%%%%%%%%%%%%%%%%%%%%%%%

\subsection{Non-Boussinesq effects}

One possible measure of Boussinesq conditions is that the fractional
change in density across the layer,
\begin{equation}
\frac{\Delta \rho}{\rho} = \alpha \Delta T,
\end{equation}
must be small. On the basis of a comparison to the Boussinesq
problem at the onset of convection, it is generally assumed that
values of $\alpha \Delta T < 0.2$, or a 20\% variation of density
across the flow thickness, is acceptable. This criterion is indeed
satisfied up to very high values of $Ra$ (above $10^{15}$ for one
set of data \cite{NandS_JFM} and above $10^{16}$ for another
\cite{natu}), although there is no assurance that asymmetries of
this magnitude are irrelevant at such high $Ra$. In fact, a more
stringent requirement by a factor of 4 was adopted in Ref.\
\cite{NandS_JFM}.

Because of the importance of the non-Boussinesq effects, as
discussed in Ref.\ \cite{NandS_JFM}, recent attention has been
focused on them. The earliest exploration was by Wu \& Libchaber
\cite{libc91}, who reported top-bottom asymmetry in boundary layers
as a main characteristic and a drop in the ratio of temperature drop
across top to bottom boundary layer as the Rayleigh number
increases. Velocity profiles measured in a follow-up paper
\cite{libc98}, at lower $Ra$, using glycerol, also showed an
asymmetry. Ahlers and collaborators \cite{ahle06} showed that the
non-Boussinesq effects depend on the fluid, as one could expect. For
water, $Nu$ showed a modest decrease with increase in $\Delta T$.
For ethane, they found larger $Nu$ than in the Boussinesq case,
nearly $10\%$ higher when $\alpha \Delta T=0.2$.

Because there are many possible non-Boussinesq effects and their
relative importance depends on the fluid and the operating
conditions, it is difficult to study these effects systematically in
experiments. A numerical computation by \cite{suki07} in two
dimensions, with glycerol as working fluid, showed that effects on
$Nu$ were marginal, with some decrease in $Nu$ with $\alpha \Delta
T$ for $Ra>10^7$. In \cite{same}, these effects have been explored
in three dimensional convection, also computationally. The finding
is that---at least for conditions corresponding to cryogenic helium
gas at modest Rayleigh numbers---while viscosity plays an important
role in diminishing the movement of plumes to the interior of
convection it is the coefficient of thermal expansion that affects
heat transport most.

\subsection{Whither helium experiments?}

While thermal convection has been studied for quite some time, the
recent surge of interest has been triggered by helium
experiments---even in theory and simulations. Indeed, experiments
were ahead of theory and simulations about two decades ago. Since
then, theory has been making its presence felt slowly and
simulations have been making considerable inroads. Experiments have
surely extended the parameter ranges, but, just as surely, they have
not kept up the pace of sophistication. A major step in the
understanding of the problem will occur only if a major improvement
in experimental sophistication takes place. It is therefore useful
to take stock of the situation briefly. It is perhaps useful even to
raise the question as to whether the promise of helium is realizable
in its entirety anytime soon.

It has been recognized abundantly that the problem is with
instrumentation and with probes of the desired temporal and spatial
resolution. It is not clear to us that smaller probes based on the
principles of standard thermal anemometry are the solution to the
problem, part of which arises because the use of helium raises the
Reynolds number of the probe itself to a higher value than in
conventional fluids, leading to unfavorable (and poorly analyzed)
heat transfer characteristics.

In thermal convection flows, where some direct knowledge of the
velocity would be most desirable even at scales much larger than the
Kolmogorov length, the use of hot and cold wires is further
complicated by the fact that they require a steady flow---and the
mean wind is effective only near the boundaries and also becomes
weaker with $Ra$. Complications arise from the simultaneous presence
of temperature fluctuations in the cell and the temperature
fluctuations of the wire due the velocity fluctuation that is
intended to be measured.

Even if single-point measurements were possible successfully, the
need to measure the entire velocity field in a turbulent flow
remains to be addressed. While a number of hot wires at several
points can be used to obtain some spatial information, there is a
limit to this procedure. In principle, we may use the Particle Image
Velocimetry (PIV) to obtain an entire two-dimensional section of the
turbulent flow field at a given instant in time. In fact, PIV has
been applied recently to liquid helium grid turbulence at 4.2K
\cite{whit,QFSPIV}, in counterflow turbulence \cite{vans} and in
helium II turbulence \cite{bewl}. It should be pointed out that
because of the enormous amount of information in any one image, it
is difficult to process very high data rates from this type of
measurement. Consequently, time evolution of the flow cannot be
easily obtained.

Particle selection and injection remain the fundamental hurdle for
PIV measurements at low temperatures. Liquid helium has a relatively
low density, and this makes it harder to find suitably buoyant
particles that are also not too large. The use of hydrogen particles
that match the density of helium has been the most promising step in
this direction \cite{bewl}, but better control of the particle
generation is needed to render the technique routinely usable. It is
equally important to better understand the interaction of particles
with the mixture of normal and super fluids \cite{carl}.

The seeding of helium gas for thermal convection experiments is
probably even more difficult owing to the large variation of the
density, and its nominally small value, which at best is less than
half that of the liquid phase. However, the liquid, as mentioned
above, can also be used to attain high $Ra$, though at the expense
of a large range. The compensation is that we know that the liquid
flow can be seeded to some level of adequacy.

Flow visualization can focus experimental---and even
theoretical---efforts, and yet this domain has not been well
developed for cryogenic helium. We believe that there is a huge
pay-off because most existing flow visualizations in water and other
room-temperature fluids are at low to moderate Rayleigh numbers, and
the intuition that one derives from low $Ra$ cannot easily be
extended to high $Ra$. There are no technological barriers to
perfecting the present efforts---only one of integrating various
components together. We may also remark that it is not easy to test
new particles in the actual low temperature environment. In
experimental phase, White et al.\ \cite{whit} had resorted to
testing in a pressurized $SF_6$ environment, where the density could
be matched to that of liquid helium.

Where density gradients exist in the flow, visualization can occur
in the absence of tracer particles, using shadowgraphs (which
depends on the density gradient) or schlieren technique (which
depends on the second derivative of the density). It has been
demonstrated \cite{luca2} that shadowgraphy can be used in helium I
to visualize even weak flows near the convective onset. A light beam
reflected from the cell displays intensity variations resulting from
convergence or divergence as a result of gradients in the refractive
index. In the case of thermal convection, these indicate the average
temperature field. Note that the technique does not give local
information, but can be used to visualize only global flows. In the
case of large apparatus, installing an optically transparent but
thermally conducting plates is a non-trivial task.

For the case of turbulence under isothermal conditions, it would be
possible to use helium 3 as a ``dye marker" for shadowgraphs.

Scattering of ultrasound is another method that can in principle be
used for velocity measurements in helium. It can be used in the gas
phase which makes it a plausible candidate for cryogenic convection
experiments. However, there would be substantial problems with
achieving sufficiently high signal-to-noise ratio resulting from a
mismatch of acoustic impedance between the sound transducers and the
helium. The work in this direction \cite{stei,baud} has not yet been
adopted in cryogenic helium.

In summary, one part of the promise of helium (namely large values
and ranges of the control parameters) has been amply established;
flows with huge values of $Ra$ and $Re$ have indeed been generated
in laboratory-sized apparatus. However, the second part of the
promise (of being able to develop versatile techniques for precise
measurements of velocity and vorticity) has lagged behind
substantially, despite some impressive efforts. This is the aspect
that needs substantial investment.

A speculative possibility in this direction, cited in \cite{jltp},
is the laser-induced fluorescence of metastable helium molecules,
which can be the ``tracer" particles in PIV measurements. This could
be a sensitive tool with good spatial resolution (only molecules in
the intersecting region of two crossed lasers would be excited to
metastable states), and it is also possible that a single molecule
could be detected spectroscopically. Above 1K, the molecules would
likely move with the normal fluid component, allowing its velocity
profile to be detected, provided that the molecules are not trapped
on vortex lines. Such a technique would be complementary to more
typical second sound attenuation measurements.

Once the instrumentation issues are clearer, we need to seriously
consider an experiment that can combine moderate aspect ratio (say,
4) with high $Ra$, constant $Pr$, and Boussinesq conditions. Such an
experiment is probably not without considerable technical
difficulties. A large scale low temperature apparatus could be
constructed, say at a facility like CERN or BNL, where there is an
adequate refrigeration capacity. Having a horizontal dimension of,
say, 5 meters or more would probably require some type of
segmentation of the plates with multiplexing of the heating and
temperature control. Fundamentally, this is no more complicated than
the mirror arrays used in astrophysical observation.  The bottom
plate, which has a constant heat flux condition imposed, can be
arbitrarily thick since it can be supported from below. The top,
temperature controlled plate would probably have some limitations in
this respect. Estimates have been made of the cooling power required
for a cell that is 5 m in diameter but also 10 m tall, and it is
around 200W or less, which is not a severe requirement.

\section{Superfluid turbulence}
\label{superfluid}

We now review aspects of liquid helium below the lambda point,
called helium II. At low velocities, helium II flows without
friction but the situation changes when velocity exceeds a critical
value: the fluid enters a state in which quantized vortices are
formed spontaneously into a self-sustained tangle---except under
controlled conditions such as rotation, for which the vortices are
all aligned with the direction of rotation. The vortex lines move
about in the background of freely moving elementary excitations,
which one may regard as something of background ``gas particles".
The vortices scatter the excitations as long as there is relative
velocity between them and the ``particles", thus generating
friction. The vortices are estimated to be of the order of an
angstrom in diameter, and, as was recognized by Onsager in 1949,
quantum mechanics constrains the circulation around them to be $n
\kappa/m$, where $\kappa$ is the Planck's constant and $m$ is the
mass of the helium atom; the integer $n=1$ normally. However, the
irrotational flow further away from the core of the vortices is
classical. The motion produced by a vortex tangle can be quite
complex because of its complex geometry and is called superfluid
turbulence.

\subsection{Analogy to classical turbulence}

One of the recent findings \cite{tabe} is that the superfluid
turbulence has the Kolmogorov form for the spectral density with a
well-defined $-5/3$ power, independent of whether the fraction of
the normal fluid (corresponding to excitations) is negligible or
dominant. This result may not seem surprising if one takes the stand
that any nonlinearly interacting system of many scales will behave
similarly to the classical Kolmogorov turbulence in the inertial
range \cite{monin}. What is needed are the mechanisms of excitation
at some large scale and dissipation at the small scale, with no
further detail mattering in the inertial range. However, several
problems come to the fore upon closer scrutiny.

First the dissipation mechanism: It is generally accepted that the
short wavelength Kelvin waves are responsible for dissipation
\cite{vinePRB}. These waves are created presumably by the impulse
associated with the reconnection of vortices. For temperatures of 1
K and above, Kelvin waves are damped out by the background
excitations (the normal fluid) thus providing the dissipation
mechanism. For lower temperatures, for which the normal fluid is
negligible, the energy is radiated away as sound at sufficiently
small wavelengths. For radiation to be effective, one needs high
velocities and short wavelengths: modest motion of vortices will not
do. Higher velocities are possible very close to the vortex core
because of the inverse power law of the potential velocity
field---and also because of reconnection events, which produce
cusp-like local structures with sharply repelling velocities.

Regarding the forcing scale, in experiments with a pull through grid
in helium II \cite{stal}, it is conceivable that the forcing is
produced very similarly to that in classical turbulence, and is
related to the mesh length and the time of evolution of the
turbulence. In some simulations, the forcing scale cannot be defined
unambiguously. For instance, in the important foray into superfluid
turbulence that was made by Schwarz \cite{schw}, it appears that
forcing scale was the size of the computational box, as in the case
of the simulations of the Taylor-Green problem by Nore et al.\
\cite{nore} and Araki et al.\ \cite{arak}. However, it also appears
that reconnections play an important role in determining this scale.

As another perspective on the same issue, the occurrence of the
$-5/3$ spectrum in superfluid turbulence may be regarded as
surprising if one takes the stand that the key mechanism for energy
transfer across scales in hydrodynamic turbulence, namely vortex
stretching, is absent in superfluid turbulence: no intensification
and break-up due to vortex stretching is possible. It is the vortex
break-up due to reconnections, not vortex stretching, that appears
to be the key to the spectral distribution here. If this is true, it
is an interesting to speculate about the central importance attached
to vortex stretching in classical turbulence.

To be sure, one should look closely at the veracity of claims about
the $-5/3$ power law. Our view is that the available evidence is too
fragile to sustain the claim on the existence of the $-5/3$ spectrum
in experiment or simulations. In experiments, the only real piece of
evidence comes from Ref.\ \cite{tabe}, but at least to us it is not
exactly clear what is being measured at the low end of the
temperature (below 1 K), despite a nice assessment in Ref.\ \cite
{VN}. In simulations of superfluid turbulence, the result is
unconvincing because the computational box size is still a long way
from reaching the size characteristic of those in classical
turbulence. We make a strong case for such an effort.

At slightly higher temperatures than 1 K, the available evidence is
also due to \cite{tabe}. The data concern a mixture of superfluid
and normal helium and it is hard to disentangle the two. The
measurements of \cite{stal}, though intrinsically exciting in
addition to having instigated the recent recent interest in the
problem, are indirect. Here, one measures the decay of superfluid
vorticity (with its own caveats) and notes that the behavior is
similar to that of the classical vorticity. From this one can
compute the energy dissipation rate and infer the classical
Kolmogorov spectrum.

Our conclusion is $not$ that the $-5/3$ power is ruled out, but that
the evidence is soft at present; one needs to produce more direct
and convincing evidence.

There is another interesting wrinkle. If one assumes that the
wavelength of the Kelvin waves which dissipate or radiate the energy
are very small compared to the Kolmogorov scale, it is plausible to
infer the spectral amplitude of fluctuations of superfluid velocity
in the intermediate region between Kolmogorov and Kelvin scales.
Presumably, the only relevant parameter in that range is the strain
rate at the Kolmogorov scale, quite like the situation of the
passive scalar spectrum at high Schmidt numbers. It then follows
from dimensional reasoning that one should expect a $-1$ power for
the spectrum in that region. Although no one has measured the
spectrum directly, indications from the decay of superfluid
vorticity are that the energy spectrum is consistent with a $-3$
power law \cite{skrb}. This behavior is poorly understood at
present.

\subsection{Visualization of quantized vortex lines}

One of the exciting developments of recent few years is the
visualization of quantized vortices and their reconnection using
small neutral particles \cite{bewl,bewl1}. These particles are made
by the {\em in situ} freezing of mixtures of hydrogen and helium.
While these visualization studies have confirmed some interesting
aspects of quantized vortices such as rings and reconnections, the
particles are still too large compared with the diameter of the
vortices (by a factor of about $10^4$). Thus, while it is easy to
convince oneself that the particles get attracted to vortex cores
and decorate them, it is obvious that the particles are not always
passive. One can calculate conditions under which the inertia of the
particles has marginal influence on vortex lines, but there is no
controlled means to ensure that this happens always: One would have
to devise smaller particles before one can be confident of the fine
details.

\subsection{Concluding remarks on superfluid turbulence}

At least in the initial stages when the study of superfluid
turbulence was brought closer to classical turbulence community, one
of the hopes was that one might be able to create enormous Reynolds
numbers in modest-sized facilities using helium II. However, this
has turned out to be impossible in principle (though, to avoid
confusion, we should reiterate that this goal has been successfully
reached with helium I and the gaseous phase). The difficulty with
helium II is that the superfluid vorticity introduces an effective
kinematic viscosity which is of the same order as the kinematic
viscosity of helium I \cite{stal,VN}. While there is a lot to learn
and understand about superfluid turbulence as a subject of intrinsic
interest, it is unlikely that it will unlock new paradigm for
understanding classical turbulence.

The new directions of superfluid turbulence concern helium 3 at much
colder temperatures.

\section{Final Remarks}

If we are interested in discovering laws underlying systems with
many strongly interacting degrees of freedom and are far from
equilibrium, it is important to begin with a study a few of them
with the same rigor and control for which particle physics, say, is
well known. We can probably make the case that hydrodynamic
turbulence, which arises in flowing fluids, is an ideal paradigm.
Our first point is that the dynamical equations for the motion of
fluids are known to great accuracy, which means that understanding
their analytic structure can greatly supplement experimental
queries; in just the same way, computer simulations---even if they
require much investment of time and money---can be far more useful
here than for many other problems of the condensed phase, in which
the interaction potential among microscopic parts is often simply an
educated guess. The stochasticity of turbulence (and of all systems
that are driven hard) means that one may discern only laws that
concern statistical behavior. If we are fortunate, these laws are
universal in some well-understood sense. This is the way we regard
the ``problem of turbulence".

While we have not yet reached a state when we can declare victory
(perhaps that may never happen in a strict sense), the ``problem of
turbulence" is being slowly chipped away by understanding, albeit
partially, its several aspects. This review has touched a few
aspects of the problem in which considerable progress has been made
recently. There is, of course, much to do, and one needs to
understand both the richness of the problem and the discipline
needed to make a dent in one of its non-trivial aspects.

\textbf{Acknowledgements:} This work was supported in part by the
US-Israel Binational Science Foundation.


\begin{thebibliography}{99}
\bibitem{footnote}
We chose the form of $\ell^+(\zeta)$ such that near the wall such
that its expansion does not have a term quadratic in $\~\zeta$, in
agreement with the DNS data.
\end{thebibliography}
\end{document}